\documentclass{aa}  

\usepackage{xcolor} 
\usepackage{txfonts}
\usepackage{scalefnt}
\usepackage{pdflscape}
\usepackage{enumerate}
\usepackage{float}
\usepackage{graphicx}
\usepackage[caption=false]{subfig}
\usepackage{rotating}
\usepackage{txfonts}
\begin{document} 

\newbox\grsign \setbox\grsign=\hbox{$>$} \newdimen\grdimen \grdimen=\ht\grsign
\newbox\simlessbox \newbox\simgreatbox
\setbox\simgreatbox=\hbox{\raise.5ex\hbox{$>$}\llap
     {\lower.5ex\hbox{$\sim$}}}\ht1=\grdimen\dp1=0pt
\setbox\simlessbox=\hbox{\raise.5ex\hbox{$<$}\llap
     {\lower.5ex\hbox{$\sim$}}}\ht2=\grdimen\dp2=0pt
\def\simgreat{\mathrel{\copy\simgreatbox}}
\def\simless{\mathrel{\copy\simlessbox}}
\newbox\simppropto
\setbox\simppropto=\hbox{\raise.5ex\hbox{$\sim$}\llap
     {\lower.5ex\hbox{$\propto$}}}\ht2=\grdimen\dp2=0pt
\def\simpropto{\mathrel{\copy\simppropto}}

\title{Chemical evolution of Na, Mg, and Al in the Galactic bulge from UVES data }
\author{
R. P. Nunes\inst{1}
\and
B. Barbuy\inst{1}
\and
C. Chiappini\inst{2}
\and
A. Fria\c ca\inst{1}
\and
M. C. Jelin\inst{1}
\and
V. Hill\inst{3}
\and
M. Zoccali\inst{4}
\and
A. Renzini\inst{5}
\and
A. P\'erez-Villegas\inst{6}
\and
H. Ernandes\inst{7}
\and
S. O. Souza\inst{8}
\and
M. J. Castro\inst{9,10}
}
\offprints{B. Barbuy}
\institute{
Universidade de S\~ao Paulo, IAG, Rua do Mat\~ao 1226,
Cidade Universit\'aria, S\~ao Paulo 05508-900, Brazil
\and
Astrophysikalisches Institut Potsdam, An der Sternwarte 16, Potsdam, 14482, Germany
\and
Universit\'e de Sophia-Antipolis,
 Observatoire de la C\^ote d'Azur, CNRS UMR 6202, BP4229, 06304 Nice Cedex 4, France
 \and
 Universidad Catolica de Chile, Departamento de Astronomia y Astrofisica, Casilla 306, Santiago 22, Chile
 \and
INAF-Osservatorio Astronomico di Padova, Vicolo dell'Osservatorio, 5, 35122 Padova PD, Italy
\and
Instituto de Astronom\'ia, Universidad Nacional Aut\'onoma de M\'exico, A. P. 106, C.P. 22800, Ensenada, B. C., M\'exico
\and 
Nicolaus Copernicus Astronomical Center, Polish Academy of Sciences, ul. Bartycka 18, 00-716 Warsaw, Poland
\and
Max Planck Institute for Astronomy, K\"onigstuhl 17, D-69117 Heidelberg, Germany
\and
Landessternwarte, Zentrum f\"ur Astronomie der Universit\"at Heidelberg, Königstuhl 12, D-69117 Heidelberg, Germany
\and
Fakult\"at f\"ur Physik und Astronomie, Universit\"at Heidelberg, Im Neuenheimer Feld 226, D-69120 Heidelberg, Germany
}

   \date{}

 
\abstract
  {The formation of the Galactic bulge remains incompletely understood, with evidence pointing to different stellar populations, including a bar-driven component, an inner-disk population, and an older spheroidal component. Chemical abundances provide critical constraints on the origin of these populations, particularly for odd-Z elements such as Na and Al, as well as Mg, whose behaviour at high metallicity is still a matter of debate.}
      {We aim to investigate the presence of overabundances of Na, Mg, and Al in Galactic bulge stars, with particular emphasis on the metal-rich regime, and to evaluate their consistency with predictions from chemical evolution models.}
   {We re-derived the abundances of Na, Mg, and Al for a sample of 55 bulge red giants previously analysed in the literature. Our study is based on high-resolution UVES spectra obtained with the ESO Very Large Telescope and employs spectrum synthesis using the \textsc{Turbospectrum} code, with updated atomic and molecular line lists. We also re-derived the abundances of C, N, and O.}
   {We find somewhat lower abundances of Mg and Al at the metal-rich end than previous studies, while a fraction of the metal-rich stars still exhibit significant Na enhancements. These enhancements persist when different sets of stellar parameters are adopted, indicating that they are robust. Mg shows smoother trends with metallicity, broadly consistent with expectations.}
   {The presence of Na-enhanced stars at high metallicity is difficult to reconcile with standard chemical evolution models and suggests additional enrichment processes in the bulge, or a particular behaviour of stellar yields with metallicity. The Na enhancement could be due to metallicity-dependent yields from massive stars, not taken into account in available models, and/or enrichment by asymptotic giant branch stars, or due to second-generation stars evaporated from globular clusters, the latter option
arising because for the metal-rich ([Fe/H]$>$0) stars a Na-O anti-correlation appears to occur.}


  \titlerunning{Chemical evolution of Na, Mg, and Al in the Galactic bulge from UVES data}
   \maketitle
%

\section{Introduction}

Hierarchical scenarios of galaxy formation \citep{whiterees78,whitefrenk91} predict 
that the central regions of galaxies formed early and rapidly. Observations indicate that galaxies grow inside-out \citep[e.g.][]{conselice14}, 
implying that a large fraction of the oldest
stars in the Milky Way formed early and are now concentrated in the bulge 
\citep[e.g.][]{gao10,tumlinson10,barbuy18a}.

The bulk of bulge stars has an estimated age
of about 11 Gyr \citep[e.g.][]{zoccali03,clarkson08,valenti13,renzini18,joyce23}, 
while its oldest component, commonly associated with a spheroidal bulge, is inferred
to be as old as $\sim$13 Gyr as seen from ages of bulge
globular clusters with moderate metallicities
\citep[e.g.][]{souza24,ortolani26}. The fraction of intermediate-age
stars instead remains unsettled \citep{bensby17,renzini18,joyce23}.

The inner Galaxy later evolved through bar-driven secular processes, 
forming a boxy/peanut-shaped bulge on still debated timescales of 3 \citep{nepal24} to 8 Gyr \citep{buck18,bovy18,sanders24}.
In addition to the dominant bar component, observations also indicate the presence of an older spheroidal population. 
However, the detailed processes of the Galactic bulge formation remains unresolved.

Previous work has established that the Galactic bulge hosts multiple stellar populations, including a metal-poor component. \cite{zoccali17} identified substructure in the bulge metallicity distribution function (MDF) consistent with distinct populations, while \cite{lim21} showed, using 
Blanco DECam Bulge Survey  \citep[BDBS;][]{johnson20} red clump stars, that the bulge can be described as a combination of a metal-poor spheroid and a more metal-rich boxy/peanut component. A spectroscopic MDF decomposition by \cite{queiroz21} recovered multiple components via an orbital-chemical analysis, though with a more limited sensitivity to the most metal-poor regime as the dataset was based on 
the Apache Point Observatory Galactic Evolution Experiment (APOGEE; \citealt{majewski17}). In this context, \cite{nepal26}  identified a spheroidal bulge component peaking at [Fe/H] $\simeq -$0.7 dex through detailed stellar population disentanglement, combining APOGEE and RVS-CNN datasets \citep{guiglion24}.
\citet{zoccali17} estimated a fraction of 50\% of bulge stars belonging to the bar, and the other half grossly being a spheroidal bulge. The component that \citet{nepal26} identified as a pressure-supported spheroidal bulge would represent no more than 20\% of stars, and this is the component in which we are interested. 
The present-day bulge therefore comprises multiple overlapping components, including bar-driven populations, a possible primordial spheroid, and contributions from thin and thick disks, the inner halo, and accreted systems \citep{bland-hawthorn16,queiroz21,ruiz-lara26}.
Disrupted GCs may contribute to this mix, being indicated particularly through stars with multiple-population abundance patterns \citep{schiavon17,fernandez-trincado17,souza24b}.

Chemical abundances provide key constraints on these populations. Alpha elements such as O and Mg are produced in massive stars \citep{woosley95} and are enhanced in old populations formed on short timescales, reflecting enrichment by Type II supernovae \citep[e.g.][]{mcwilliam16,friaca17,barbuy18a,matteucci21}. In contrast, the odd-Z elements Na and Al show a more complex behaviour, with yields depending on metallicity and neutron excess (Woosley, private communication).
Given the expected contribution from GC debris, multiple stellar populations in GCs further complicate this picture, as second-generation stars show enhancements in N, Na, and Al and depletion in C and O \citep{carretta09a,carretta09b,gratton12,piotto15,renzini15,milone17,bastian18,milonemarino22}.
Indeed, a fraction of these stars may have been lost to the field  \citep{baumgardt03,weatherford23,souza24b}, which may help account for abundance patterns observed in the bulge, particularly at high metallicity.

 In recent years we have studied two samples of dynamically selected spheroidal bulge candidate stars with data in the H band from APOGEE, for [Fe/H] $< -$0.8
 \citep{razera22,barbuy23,barbuy24,barbuy25}, and [Fe/H] $>$ $-$0.8 \citep{ernandes26}.
  To further clarify the origin of the observed abundance patterns and their relation to GC debris and bulge formation processes, we now revisit a well-studied sample of bulge red giants \citep{zoccali06,lecureur07,hill11} observed at high resolution with the UVES spectrograph (R$\sim$45,000), providing another  excellent benchmark for a homogeneous reanalysis.
 This is a reference sample with data in the optical, in the sense that it is confirmed as
located in and with orbital characteristics of the Galactic bulge (to be presented in Chiappini et al., in preparation), and for which we have high-resolution UVES spectra.
Given that for this sample reliable abundances of several
elements are derived, these chemical abundances are a reference for the old spheroidal Galactic bulge.
 The main goal of this work is to reassess the stellar parameters and abundances of C, N, O, Mg, Na, and Al, focusing on the metal-rich regime, in order to verify the presence and origin of the reported Na overabundances. The chemical abundances recomputed in the present work use a different spectrum synthesis code, with respect to previous work, adopting \textsc{Turbospectrum} \citep{alvarez98,plez12}, together with updated atomic and molecular line lists, and improved molecular dissociation equilibrium data. In Paper II of this series the kinematics and orbital properties of the sample based on Gaia DR3 will be presented, with a focus on disentangling the different stellar populations.

Previous studies of bulge red giants have reported unexpectedly high Na abundances, especially at the metal-rich end. 
\citet{lecureur07} found Na enhancements that are difficult to reconcile with standard chemical evolution models, a result later noted in subsequent analyses \citep[e.g.][]{barbuy23} and also hinted at in Gaia-ESO data \citep{smiljanic09} and thick disk stars \citep{owusu26}. These findings raise the question whether such Na excesses reflect nucleosynthetic effects, GC debris, or limitations in abundance analyses. In addition, precise abundances of Mg and Na  also play a crucial role as these can be used as indicators of an in situ or ex situ origin in the Galaxy, as demonstrated by
\citet{hawkins15,nissen10}, and \citet{nissen24}.

The manuscript is organized as follows.
In Sect. \ref{sample} the observations are described. 
In Sect. \ref{calculations} 
 the basic stellar parameters are listed, as are atomic constants for the studied lines,
 and the calculations are described.
 In Sect. \ref{results} the results are compared with
 chemical evolution models, and discussed
 in Sect. \ref{discussion}. Conclusions are drawn in Sect. \ref{conclusions}.


\section{Observations}\label{sample}

We analysed a subset of the literature sample described in
\citet{zoccali06}, \citet{lecureur07}, \citet{zoccali08}, and \citet{hill11}
 obtained with the FLAMES-UVES spectrograph \citet{dekker00} 
at the 8.2 m Kueyen Very Large Tele- scope (VLT) at the
Paranal Observatory of the European Southern Observatory (ESO), through
ESO programmes 71.B-0617A, 73.B-0074A (PI: A. Renzini), and 71.B-0196
(PI: V. Hill).

{\it Red giants:} The present data consist of high-resolution spectra 
of 43 bulge red giants,
chosen to be located one magnitude brighter than the red clump.
The stars were observed in four fields; namely, 
Baade's window (BW; $l=1.14^{\circ}$, $b=-4.2^{\circ}$), 
a field at $\rm b=-6^{\circ}$ ($l=0.2^{\circ}$, $b=-6^{\circ}$), the Blanco field ($l=0^{\circ}$, $b=-12^{\circ}$),  and a field near NGC 6553 ($l=5.2^{\circ}$, $b=-3^{\circ}$).

{\it Red clump sample:}
The 12 red clump bulge giants in BW, 
presented in \citet{zoccali06} and \citet{lecureur07}, were further analysed
in \citet{hill11}.
This subsample has previously been described in \citet{zoccali06} and
\citet{lecureur07}, who used the same spectra and reported, respectively, the oxygen abundances and the Na, Mg, and Al abundances.

The mean wavelength coverage is 4800$-$6800 {\rm \AA}.
With the UVES standard setup 580, the resolution is R $\sim$ 45 000 for a 1 arcsec slit width, given that the fibres are 1.0" wide.
The individual spectra were co-added, taking into account the
  different observation epochs using the IRAF task imcombine to eliminate
  cosmic rays. The signal-to-noise of
 our very crowded spectra is typically between 20 and 50 per
 0.017 {\rm \AA} pixel around $\lambda$6300 {\rm \AA}.



\section{Abundance analysis}\label{calculations}

\begin{figure}
\includegraphics[angle=0,width=9cm]{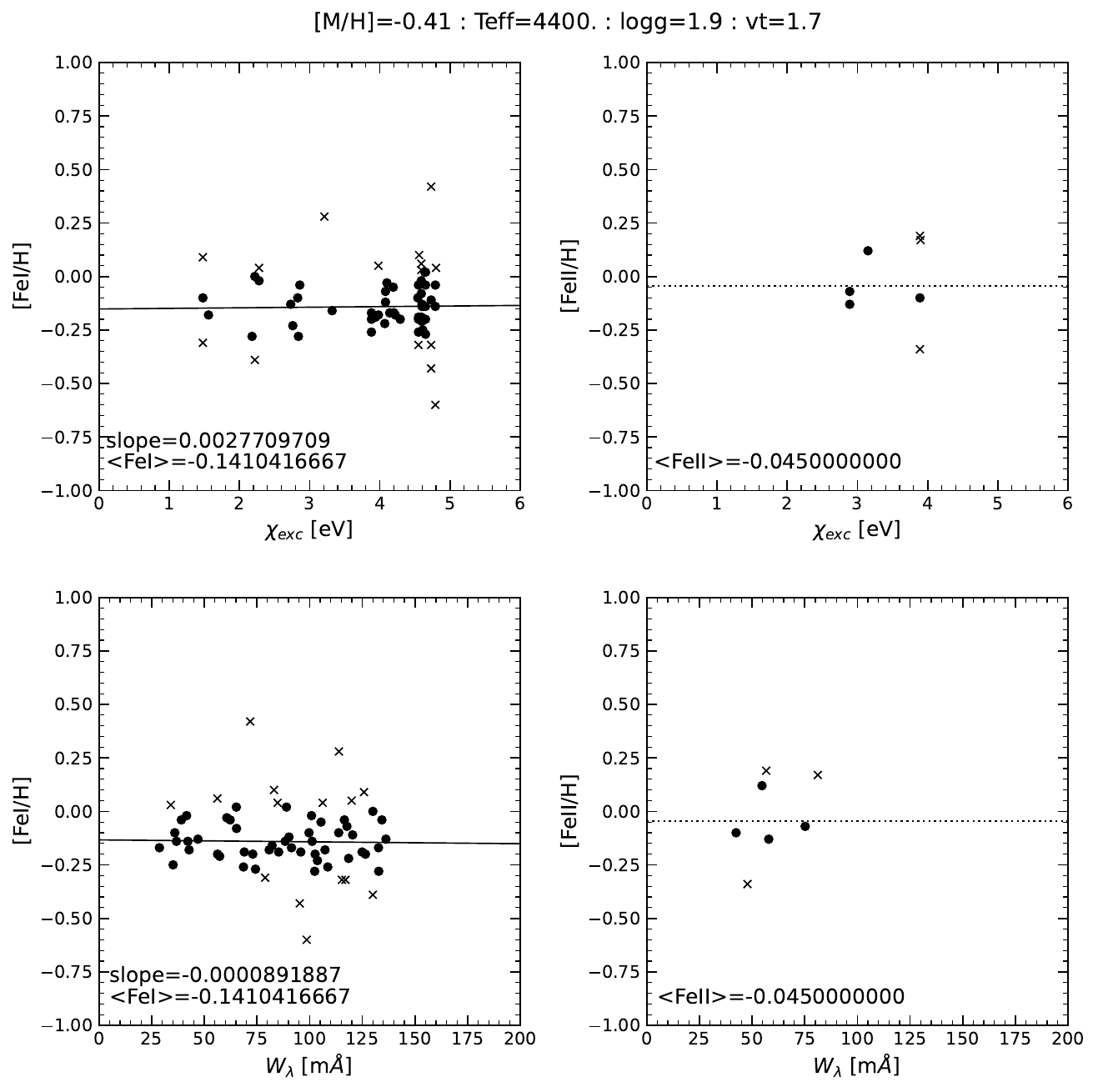}
\caption{Spectroscopic equilibrium for star B6-b4, validating the atmospheric parameters ($T_{\rm eff}=4400$~K, $\log g=1.90$, $v_{\rm t}=1.7$~km/s, $[Fe/H]=-0.41$). Points represent individual abundances for each measured iron line; filled circles denote adopted lines, while `X' symbols indicate lines
too discrepant from the mean, probably due to blending, or
to defects in the spectra such as cosmic rays or noise, and they are
excluded from the analysis. \textit{Left panels}: [Fe~I/H] abundances as a function of excitation potential ($\chi_{\mathrm{exc}}$; \textit{top}) and equivalent width ($W_{\lambda}$; \textit{bottom}). Slopes consistent with zero validate $T_{\mathrm{eff}}$ and $v_t$, respectively. \textit{Right panels}: [Fe~II/H] abundances vs $\chi_{\mathrm{exc}}$ (\textit{top}) and $W_{\lambda}$ (\textit{bottom}). The ionization equilibrium, verified by the agreement between $\langle \mathrm{Fe\,I} \rangle$ and $\langle \mathrm{Fe\,II} \rangle$, validates $\log g$.}
\label{meafs}
\end{figure}

The stellar parameters -- the 
effective temperature (T$_{\rm eff}$), surface gravity (log g), 
metallicity ([Fe/H]),
and microturbulence
velocity (v$_{\rm t}$) from \citet{zoccali06}, and \citet{lecureur07} -- were verified here
by remeasuring equivalent widths using DAOSPEC \citep{pancino08}, 
as well as the code Multiple Element Abundance Fit Software - 
MEAFS\footnote{https://github.com/MatheusJCastro/meafs},
by \citet{castro26},
and examining the excitation
and ionization equilibria, using \ion{Fe}{I} and \ion{Fe}{II} lines.
One example is shown in Fig. \ref{meafs} for star B6-b4.
This verification has shown that the stellar parameters obtained by \citet{lecureur07}
do follow the ionization and excitation equilibria.
We also tested the parameters by \citet{johnson14} and \citet{jonsson17,lomaeva19},
for the sub-samples analysed by them, and the parameters are also confirmed. 
Using these three sets of stellar parameters,
we recomputed Na, Mg, and Al, as reported in Sect. \ref{sectuncertainties}.

The calculations of synthetic spectra were carried out 
using the code \textsc{Turbospectrum} from \citet{alvarez98} and \citet{plez12}, together with
MARCS model atmospheres from \citet{gustafsson08}. 
The reference solar abundances are from \citet{asplund21}.
Elemental abundances were obtained through line-by-line spectrum synthesis calculations for the lines given in Table \ref{linelist}. 
For consistency, we re-derived the C, N, O abundances. The results are not much
different from those previously re-derived by \citet{friaca17}, and the small differences
are due to the use of \textsc{Turbospectrum} instead of PFANT \citep{barbuy18c}, where the number of molecules
and respective constants in the dissociation
equilibrium are adopted from different sources. 
The atomic line list is retrieved from the Vienna Atomic Line list \citep{ryabchikova15}.
The molecular line lists used within the Turbospectrum include
C$_2$ \citep{brooke13}, CN \citep{sneden14}, CH \citep{masseron14},
MgH \citep{skory02}, and TiO \citep{jorgensen94}.
The available features are the SWAN C$_2$(0,1) 5635.5, the CN(5,1) 6332.2 {\rm \AA},
and the forbidden [OI] 6300.311 {\rm \AA} lines; in particular, the C$_2$ feature is shallow,
and the spectra are noisy in that region.
The newly derived C, N, O, Na, Mg, and Al abundances are reported in Table \ref{atmos}. 
For Na  the results from the four Na lines are inspected, relying primarily on the \ion{Na}{I}  6160.753 {\rm \AA} line;
this is the most reliable line because the \ion{Na}{I} 5682 and 5688 {\rm \AA} have unidentified blends,
and for 20\% of the sample, to fit these lines would require an even higher Na abundance by around 0.1 dex
and \ion{Na}{I} 6154 {\rm \AA} has blends on both left and right wings, and the continuum is less clear in
a few cases. In any case, for most (80\%) of the sample stars, all 4 lines give the same result.

\begin{figure}
\includegraphics[angle=0,width=9cm]{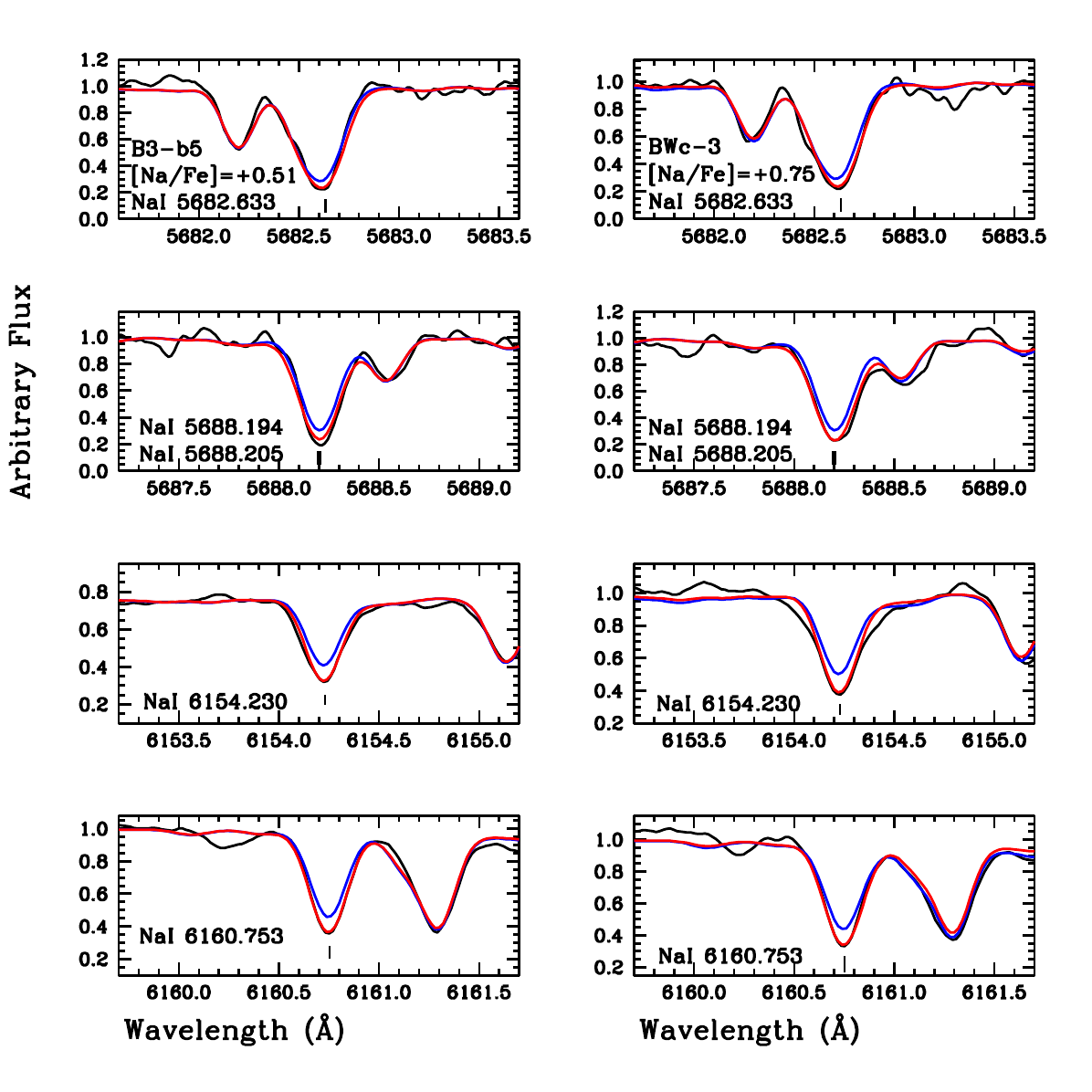}
\caption{Best fit of Na abundance from the 4 \ion{Na}{I}  lines in stars B3-b5, and BWc-3.
  The black lines represent observed spectra, the blue lines synthetic spectra computed with [Na/Fe]$=$0, and the red lines synthetic spectra computed with the final value adopted.} 
\label{figna}
\end{figure}

\begin{figure}
\includegraphics[angle=0,width=9cm]{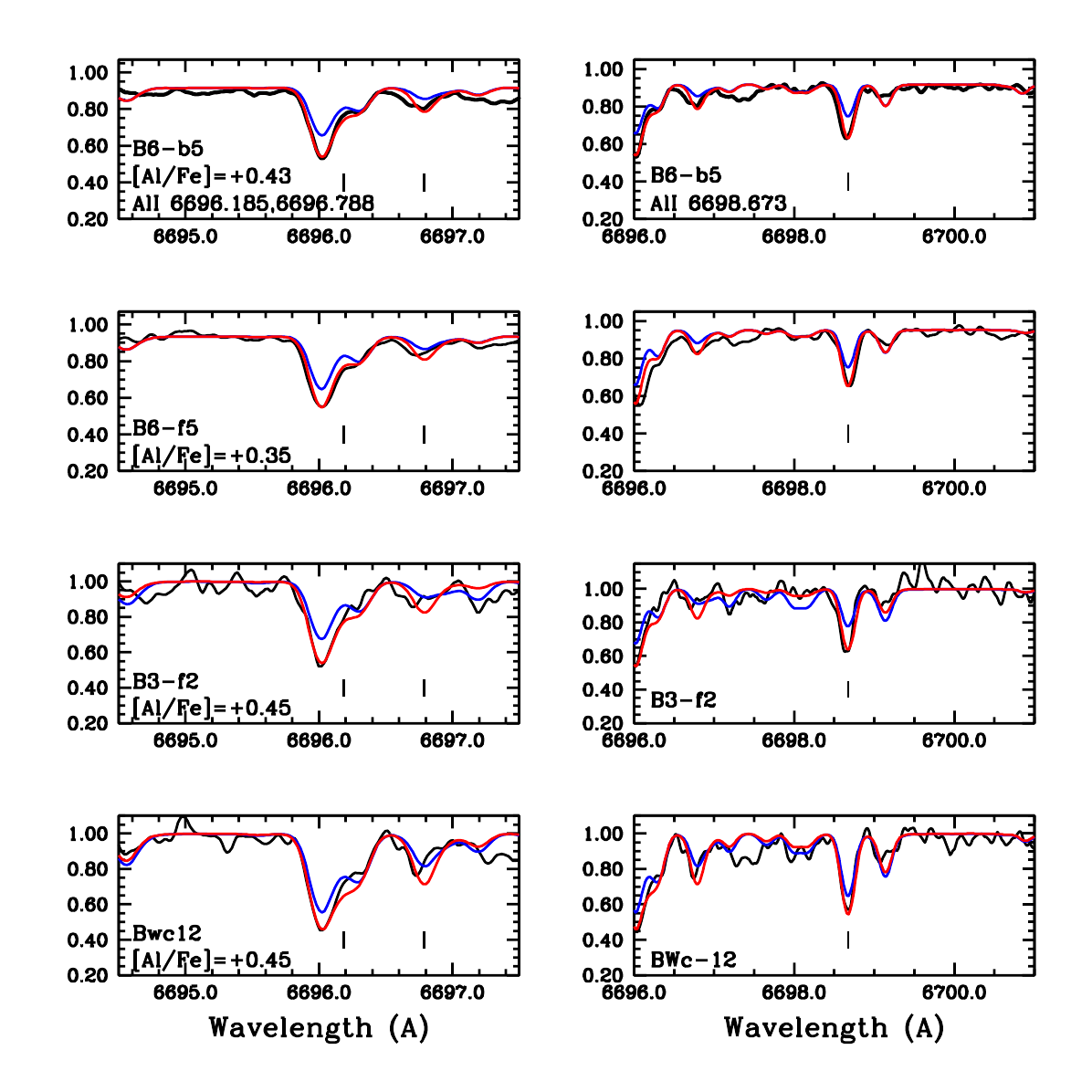}
\caption{Best fit of Al abundance from the two main \ion{Al}{I}  lines in stars B6-b5, B6-f5, B3-f2, and BWc-12.
The black lines represent observed spectra, the blue lines synthetic spectra computed with [Al/Fe]=0, and the red lines synthetic spectra computed with the final value adopted.}
\label{figal}
\end{figure}

\begin{figure}
\includegraphics[angle=0,width=9cm]{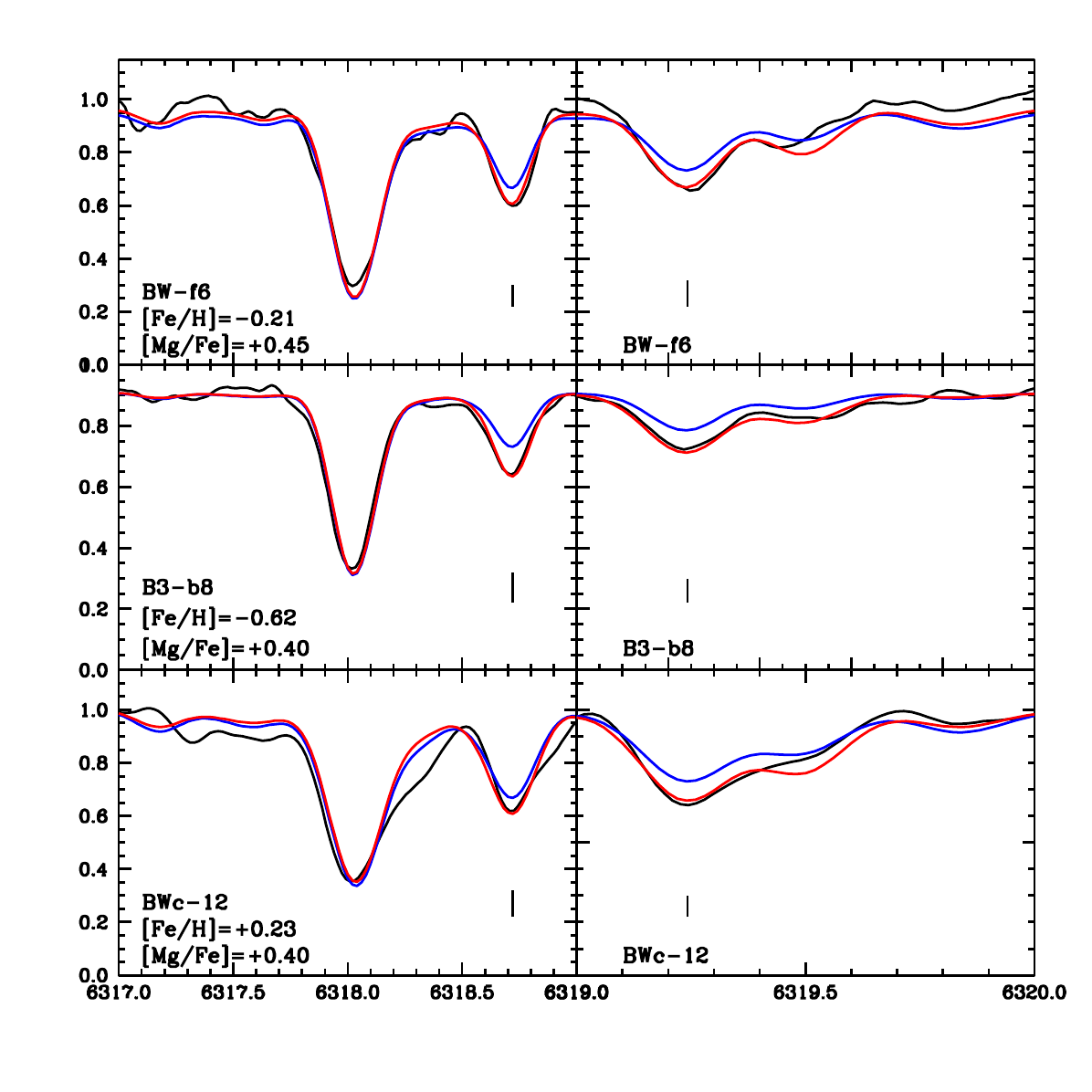}
\caption{Best fit of Mg abundance from the two main \ion{Mg}{I}  lines in stars BW-f6, B3-b8, and BWc-12. The black lines represent observed spectra, the blue lines synthetic spectra computed with [Mg/Fe]=0, and the
red lines synthetic spectra computed with the final value adopted.}
\label{figmg}
\end{figure}

The calculation of hyperfine structure was carried out for the \ion{Na}{I} 5682 and 5688 {\rm \AA } lines, using the constants
reported in Table \ref{hfs},
whereas for the 6154 and 6160 {\rm \AA } lines the splitting is already incorporated in the VALD3
line list.
Examples of line fitting are illustrated in Figs.
\ref{figna}, \ref{figal}, and \ref{figmg}.
Figure \ref{figna} shows the fits to the 4 \ion{Na}{I} lines for stars B3-b5 and BWc-3, of metallicities
of [Fe/H] = 0.11 and 0.28, respectively.
Figure \ref{figal} shows the fits to the 2 \ion{Al}{I} lines for stars B6-b5, B6-f5, B3-f2, and BWc-12.
Figure \ref{figmg} shows the fits to the 2 \ion{Mg}{I} lines for stars BW-f6, B6-f5,  B3-b8, and BWc-12.

\subsection{Other checks}

For three stars, BW-b4, BW-b5, and BW-f6, which have available observations from APOGEE
and which were reanalysed from optical and H-band spectra, in \citet{dasilva24,dasilva24b}, we
further re-derived abundances from the APOGEE spectra using the original UVES stellar
parameters from \citet{lecureur07}, as reported in Table \ref{niropt}. The results from the H band are compatible with the optical
ones. For the particular case of star OGLE-392918 = BW-f6, some discrepancy is found between
optical and H-band results. As a general comment, we can say that 
Mg and Al lines in the H band are more reliable than
the ones in the optical, whereas Na lines in the optical are clearly better than the one
in the H band, in terms of the number of lines and blending effects.
The Na line in the H band is less reliable due to blends, and also to artefacts coming
from the reductions as pointed out by \citet{hayes22}, and also shown in \citet{barbuy23}.
Another interesting issue is that a high Na abundance affects the Al abundances, indirectly
through molecule association involving several molecules containing
 Na, and Al with C, N, O: NaC, NaO, NaOH, AlO, AlO2, Al2O, and AlOH, among others,
breaking the normal balance
between molecules. Here we find a possibly spurious Na line giving
[Na/Fe]=$+$1.0 for star BW-f6, which would lead to higher Al abundances if taken into account;
however, these values would be incompatible with the results from the optical.
In conclusion, further comparisons between H-band and optical lines are of interest for the future.

\begin{table*}
\centering
\caption{Abundances for sample stars in common with APOGEE. }
\label{niropt}
\begin{tabular}{l ccc ccc ccc}
\hline\hline
\noalign{\smallskip}
Star & \multicolumn{3}{c}{545277=BW-b4} & \multicolumn{3}{c}{82760=BW-b5} & \multicolumn{3}{c}{392918=BW-f6} \\
\cline{1-4} \cline{5-7} \cline{8-10}
\noalign{\smallskip}
Element & \multicolumn{2}{c}{H band} & Opt & \multicolumn{2}{c}{H band} & Opt & \multicolumn{2}{c}{H band} & Opt \\
        & (1) & (2) & (3) & (1) & (2) & (3) & (1) & (2) & (3) \\
\noalign{\smallskip}
\hline
\noalign{\smallskip}
Na  & 0.70 & 0.65 & 0.55    &  0.37   &  0.32   & +0.27   & 0.00 & ---  & 0.00 \\
Mg  & 0.00 & 0.05 & $-$0.20 & $-$0.15 & $-$0.15 & $-$0.10 & 0.10 & 0.10 & 0.30 \\
Al  & 0.65 & 0.55 & 0.55    &  0.50  &  0.40   & +0.20 & 0.35 & 0.35 & 0.03 \\
\noalign{\smallskip}
\hline\hline
\end{tabular}
\tablefoot{Calculations were carried out with \textsc{Turbospectrum},
adopting the best-suiting stellar parameters from \citet{lecureur07}. We report previous values from
\citet{dasilva24} and present results from the H band and optical; (1) H band from \citet{dasilva24},
(2) H band from this work, and (3) optical from this work.}
\end{table*}

\begin{table}
\caption{Abundance uncertainties due to stellar parameters for the warm metal-rich star B6-f8,
and the cool metal-rich star B6-b8, for uncertainties of $\Delta$T$_{\rm eff}$ = 100 K,
$\Delta$log g = 0.2, $\Delta$v$_{\rm t}$ = 0.2 km s$^{-1}$, the
adopted error from the continuum fit uncertainty, and the
corresponding total error.}
\label{errors}
\centering
\setlength{\tabcolsep}{3.5pt} 
\begin{tabular}{lccccc}
\hline\hline
\noalign{\smallskip}
Element & $\Delta$T & $\Delta$log $g$ & \phantom{-}$\Delta$v$_{t}$ & continuum & \phantom{-}($\sum$x$^{2}$)$^{1/2}$ \\
 & 100 K & 0.2 dex & 0.2 km s$^{-1}$ & &  \\
(1) & (2) & (3) & (4) & (5) & (6)\\
\noalign{\smallskip}
\hline
\noalign{\smallskip}
\multicolumn{6}{c}{B6-f8} \\
\noalign{\smallskip}
\hline
\noalign{\smallskip}
{}[C/Fe]  & +0.00 & +0.10 & +0.00 & 0.05 & 0.11 \\
{}[N/Fe]  & +0.00 & +0.10 & +0.00 & 0.05 & 0.11 \\
{}[O/Fe]  & +0.08 & +0.13 & +0.01 & 0.05 & 0.16 \\
{}[Na/Fe] & $-$0.15 & $-$0.04 & $-$0.20 & 0.05 & 0.26 \\
{}[Mg/Fe] & +0.05 & +0.12 & +0.00 & 0.05 & 0.14 \\
{}[Al/Fe] & $-$0.10 & $-$0.01 & $-$0.10 & 0.05 & 0.15 \\
\noalign{\smallskip}
\hline
\noalign{\smallskip}
\multicolumn{6}{c}{B6-b8} \\
\noalign{\smallskip}
\hline
\noalign{\smallskip}
{}[C/Fe]  & $-$0.06 & +0.10 & $-$0.02 & 0.05 & 0.13 \\
{}[N/Fe]  & +0.20 & +0.10 & +0.03 & 0.05 & 0.23 \\
{}[O/Fe]  & +0.08 & +0.15 & +0.01 & 0.05 & 0.18 \\
{}[Na/Fe] & +0.24 & +0.12 & $-$0.02 & 0.05 & 0.27 \\
{}[Mg/Fe] & +0.05 & +0.10 & $-$0.01 & 0.05 & 0.12 \\
{}[Al/Fe] & +0.20 & +0.20 & $-$0.01 & 0.05 & 0.29 \\
\noalign{\smallskip} 
\hline 
\end{tabular}
\end{table}

\subsection{Uncertainties\label{sectuncertainties}}

Abundance uncertainties due to stellar parameters were estimated in the standard way.
Model atmosphere parameters were changed by $\pm$ 100 K in effective
temperature, $\pm$ 0.20 in surface gravity, and $\pm$ 0.20 kms$^{-1}$ in
microturbulent velocity, and continuum fit uncertainty. These uncertainties, together with continuum fit
uncertainty, were added in quadrature to estimate
a total error.
In Table \ref{errors}
we compute Na, Mg, and Al abundance unceritainties for the warm metal-rich star B6-f8
and the cool metal-rich star B6-b8.
 It can be seen that the uncertainty due to effective temperature
is non-negligible for Na in both stars, and also substantial
for Al for the cooler star.

We also computed the Na, Mg, and Al lines adopting the model atmosphere parameters derived by
\citet{johnson14}, having analysed corresponding GIRAFFE (lower-resolution fibres in FLAMES-UVES) spectra for stars in common with our sample. 
The same was done for stellar parameters re-derived by
\citet{jonsson17},  using  UVES data of 32 stars from our present sample 
\citep{zoccali06,lecureur07}.
The results are given in Table \ref{johnson}.
For Mg our abundances are compatible with those
from \citet{johnson14} and \citet{jonsson17}, as well with
our calculations with their parameters, within about $\pm$0.05.
For Na and Al, \citet{jonsson17} did not report abundances,
and our calculations with their parameters
are compatible. The comparison with \citet{johnson14}
shows that our final values are in good agreement with them,
which is the most important result, 
whereas our calculations with their parameters give
compatible Al, but
higher Na. The fact that our final abundances are in good agreement is
an indicator that the uncertainties are small.

In \citet{zoccali08} all the stars in the present sample,
also observed with GIRAFFE, were reanalysed. In a forthcoming paper,
we shall address the full GIRAFFE sample from \citet{zoccali08}.
The differences in stellar parameters 
between the original analysis by \citet{zoccali06,lecureur07} and those by \citet{jonsson17} and \citet{johnson14} were previously discussed in \citet{dasilveira18}. 
The same spectra were reanalysed by
\citet{jonsson17}, and complemented by \citet{lomaeva19}.
Their parameters
were obtained by using the software Spectroscopy Made Easy 
\citep[SME;][]{valenti96}. The SME software simultaneously
fits stellar parameters and abundances by fitting calculated
synthetic spectra to an observed spectrum.
According to the authors, the stellar parameters (T$_{\rm eff}$, log g, [Fe/H],
and v$_{\rm t}$) were derived simultaneously, using relatively weak, unblended
\ion{Fe}{I}, \ion{Fe}{II}, and \ion{Ca}{I} lines
and gravity-sensitive \ion{Ca}{I}-wings.
In the mean, the differences in parameters amount to
$\Delta$T$_{\rm eff}$\-(J\"onsson+17\--\-Zoccali+06)\-=\-$-$94 K
in effective temperatures and
 $\Delta$log~g\-(J\"onsson+17\--Zoccali+06)\-=\-+0.46 in gravities.

As regards non-local thermodynamic equilibrium (NLTE) effects, our sample of moderately metal-rich giants is expected to be affected by these effects, following the analysis in \citet{lind22}, as illustrated by benchmark stars such as Arcturus. Using grids of NLTE departure coefficients and line-by-line curves of growth, \citet{lind22} showed that both the magnitude and the sign of NLTE corrections depend strongly on line strength and wavelength. In the case of Arcturus, the corrections for Na, Mg, and Al are generally moderate, but become increasingly negative for stronger lines. This behaviour supports the applicability of similar NLTE corrections to our sample. Accordingly, 
the corrections on our sample would be 
$\Delta$A(Na)=$-$0.12, 
$\Delta$A(Mg)=$-$0.1, and
$\Delta$A(Al)=$-$0.13 dex.

Finally, to the 
abundance uncertainties due to uncertainties on
 stellar parameters, in Table \ref{errors}, we add
the uncertainty due to continuum fit, adopted as $\pm$0.05 for all analysed
elements. 
This amount was estimated by moving the continuum
up and down, showing still acceptable fits.
The final uncertainties are indicated in Figs. \ref{jonsson}, 
\ref{plotmgnaalvsfe}, \ref{plotOvsO}, and \ref{plotnaalvsmg}.

\section{Results}\label{results}

In this section we present the derived abundances of Na, Mg, and Al, compared with chemical evolution models.
Different nucleosynthesis options for explaining the metal-rich end are presented and discussed.

\subsection{Chemical evolution models}

We adopted the chemodynamical evolution models from \citet{friaca98} that were built to follow the chemical evolution in elliptical galaxies. 
The code was first applied to the Galactic bulge in \citet{friaca17}.
The bulge is assumed to be a small classical spheroid, with a baryonic mass of 2$\times$10$^9$ M$_{\odot}$
and a dark halo mass of $M_{H}$= 1.3$\times$10$^{10}$ M$_{\odot}$.
The code enables the inflow and outflow of gas, by solving the fluid equations of mass, momentum, and energy conservation, taking into account the sink terms for the gas due to star formation and source and heating terms due to the late stages of stellar evolution (supernovae, planetary nebulae, asymptotic giant branch (AGB) stars,  and stellar winds). The cooling function of the gas 
takes into account the metallicity of the gas.
These hydrodynamical equations are coupled with the chemical evolution equations.
For massive stars, we adopted the metallicity dependent yields from core-collapse SNe II from
Woosley \& Weaver (1995, hereafter WW95),
with some alterations of the yields following the suggestions of
\citet{timmes95}, 
and for metallicities of $<$ $-$2.5, 
yields from high explosion-energy hypernovae from 
\citet{nomoto13}.
Yields of SNe Ia resulting from Chandrasekhar mass white dwarfs are from \citet{iwamoto99}; namely, 
their models W7 (progenitor star of initial metallicity Z=Z$_{\odot}$)
and W70 (initial metallicity Z=0).
 We also computed models with the W7 and W70 yields of Mg and Na multiplied by 10, in an attempt to understand the SN Ia role in the behaviour of these elements at the metal-rich end of the data
(see discussions below).
Time delays for SN Ia explosions are adopted from \citet{greggiorenzini83}, with a few modifications following
\citet{mennekens10}: binaries with a total mass in the range of 3 to 16 M$_{\odot}$ for the primary, and
1.5 to 8 M$_{\odot}$ for the secondary, resulting in delays from 0.03 Gyr for the most massive binaries,
to 3.7 Gyr for the least massive ones, are considered.
Finally, yields for intermediate-mass stars ($0.8 - 8$ M$_{\odot}$) with initial Z=0.001, 0.004, 0.008, 0.02, and 0.04 are from \citet{vandenhoek97} (variable $\eta_{AGB}$ case).
A Salpeter initial mass function including stars from 0.1 to 100 M$_{\odot}$ is adopted.

Specific star formation rates (SFRs), defined as
the inverse of the timescale for the system formation, represented by
$\nu_{\rm SF}$, are given in Gigayear$^{-1}$.
In the present models we assume  $\nu_{\rm SF}=$ 3 and 1 Gyr$^{-1}$, corresponding to 
fast timescales of approximately 0.3  and 1 Gyr, respectively, for the chemical enrichment of the bulge.

We also show in the next figures, in order to allow a comparison with other chemical evolution modelling, the widely cited models by \citet{kobayashi20}, here designated K20.
It should be noted that the K20 model operates on a one-zone framework and is intended to track the chemical enrichment of the solar neighbourhood, while the present model is fully chemodynamical and designed to follow the chemical evolution of the bulge. With respect to core-collapse supernovae, the K20 model includes failed supernovae and magneto-rotational supernovae, which are not considered in our model. On the other hand, the K20 model does not include neutrino processes in low-metallicity supernovae, which provide an additional nucleosynthetic ingredient in our models.

\subsection{Results compared with chemical evolution models}

 We begin by adopting bulge chemical evolution models computed with a fairly standard set of stellar yields, as described above. As has been demonstrated in several previous studies by our group,
 the chemical evolution models applied
 to O, Mg, Si, Ca \citep{friaca17,barbuy18a,razera22}, Na, Al
 \citep{barbuy23,ernandes26},
 iron-peak elements 
 V, Cr, Mn, Co, Ni, and Cu
\citep{barbuy24,ernandes20,ernandes26}, and
P, S, K \citep{barbuy25,ernandes26}
are able to reproduce the general abundance trends observed in samples of bulge stars\footnote{See also Chiappini et al in prep (Paper II), and \citet{nepal26} for a discussion of bona fide spheroidal bulge stars and the mix of stellar populations in the bulge region.}. This provides a useful `ground-zero' baseline for our analysis.
In the present work, we focus on a direct comparison between these baseline model predictions and the observed data, now with precise chemical abundances, with the goal of identifying and interpreting any discrepancies.
Here we explore deviations between the models and the observations that are likely linked to uncertainties in stellar nucleosynthesis, concentrating on Na, Mg, and Al, but also using the revised C, N, and O abundances in the analysis. In particular, we investigate the possibility of higher yields of Na and Mg by SNe Ia  comparing the predictions from the model presented in \citet[][hereafter `standard model']{barbuy23}, with the results of the model using yields of these elements increased by a factor of 10 (the {\it NaMg-enhanced model}).

\begin{figure}
\includegraphics[angle=0,width=9cm]{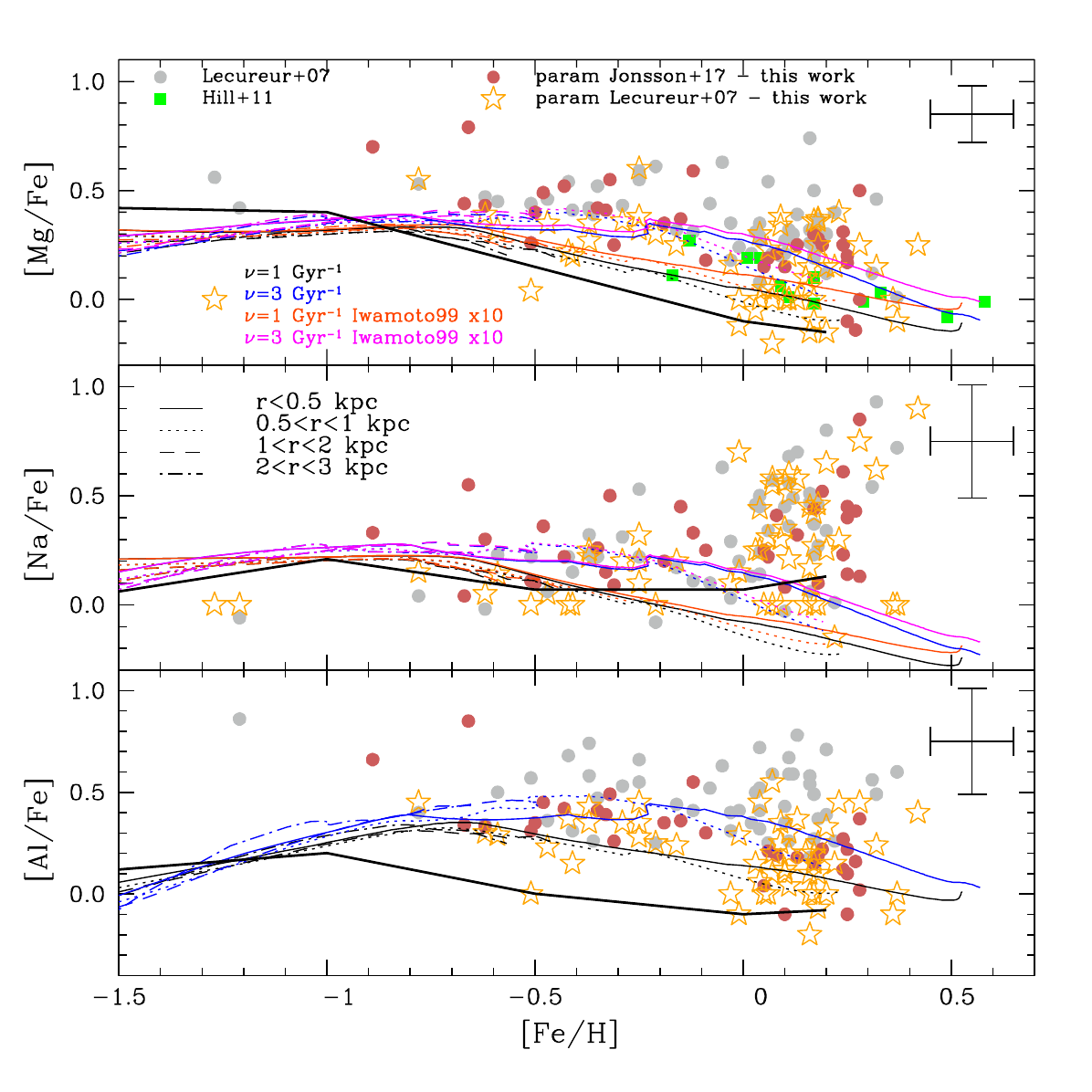}
\caption{[Na/Fe], [Mg/Fe], and [Al/Fe] vs [Fe/H]  for the present results 
compared with results from \citet{lecureur07}, and results derived using stellar parameters from \citet{jonsson17}, compared to chemical evolution models.
The orange stars represent this work and are to be compared with filled indianred circles representing this work computed with parameters from \citet{jonsson17}, filled white-grey circles \citet{lecureur07}, and filled green squares \citet{hill11}.
Chemodynamical evolution models with a specific star formation rate of
of $\nu$ = 1 (black lines) and 3 Gyr$^{-1}$ (blue lines)
that correspond to formation timescales
of 1 and 0.3 Gyr respectively, are overplotted, and these same models considering yields of SNe Ia multiplied by 10 (red and magenta lines, respectively). 
Different model lines correspond to the outputs of models computed for radii of r $<$ 0.5
(solid lines), 0.5 $<$ r $<$ 1 (dotted lines), 1 $<$ r $<$ 2 (dashed lines),  and 2 $<$ r $<$ 3 
(dash-dotted lines) kpc from the Galactic centre. 
The strong solid black lines represent the K20 model. Error bars correspond to the values in
Table \ref{errors}, adopting a mean between values for the warm and the cool stars.}
\label{jonsson}
\end{figure}

In order to show the differences between the different abundance derivations, we show in Fig.
\ref{jonsson} the Na, Mg, and Al abundances derived originally by \citet{lecureur07}, and \citet{hill11},
the present derivations using the original parameters, and the parameters by \citet{jonsson17}, and
\citet{lomaeva19} for the subsample in common with our sample.
Relative to our previous results,
the abundances derived in this work for Mg and Al are systematically lower than those by \citet{lecureur07} over the whole metallicity range. On the other hand, the intriguing Na enhancement found by them, especially for metal-rich stars, remains in the new calculations. A comparison with the original values from \citet{hill11} results in similar values for six stars and values $\sim$0.15 dex lower for the other six stars.

\begin{figure}
\centering
\includegraphics[angle=0,width=9cm]{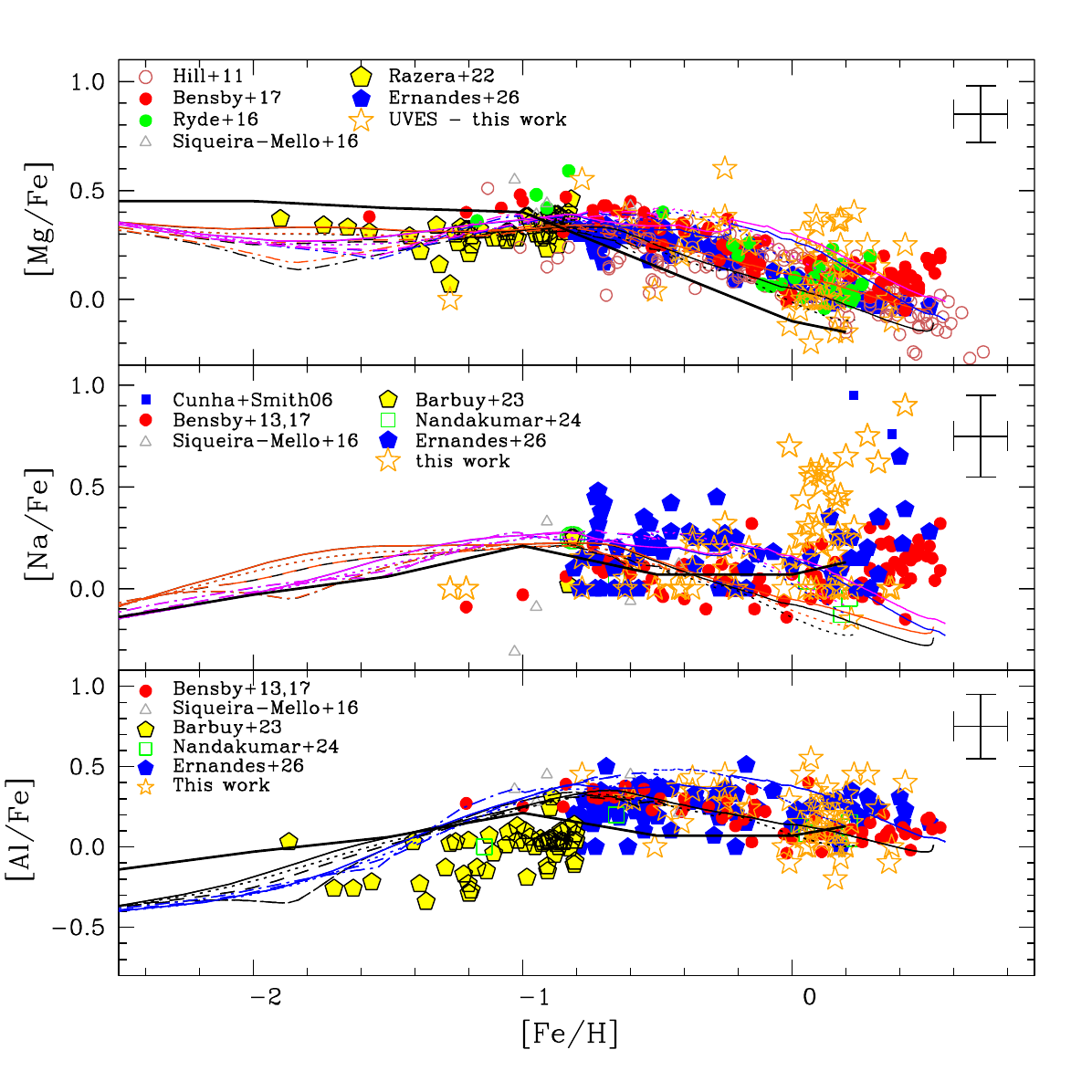}
\caption{[Mg/Fe], [Na/Fe], and [Al/Fe] vs [Fe/H]  for the present results 
compared with literature bulge samples and chemical evolution models.
The orange stars represent this work, to be compared with: filled yellow pentagons for \citet{razera22} and \citet{barbuy23}; filled blue pentagons for \citet{ernandes26}; open indianred circles for \citet{hill11};,
filled red circles for \citet{bensby17}; filled green circles for \citet{ryde16}; open grey triangles for \citet{siqueira-mello16}; open green squares for \citet{nandakumar24};
and filled blue squares for \citet{cunhasmith06};  for Na only the BAWLAS selected ones - see text.
Chemodynamical evolution models and error bars are the same as in Fig \ref{jonsson}. The K20 models are the strong solid black lines.}
\label{plotmgnaalvsfe}
\end{figure}

In Fig. \ref{plotmgnaalvsfe} we gather Mg, Na, and Al abundances for selected
samples of bulge stars, obtained from high-resolution data, and
derived with similar spectrum synthesis codes and line lists.
These include bulge stars from \citet{razera22}, \citet{barbuy23}, and
\citet{ernandes26}, studied with H-band data from APOGEE,
noting that for Na in metal-poor stars only the best spectra selected in
BACCHUS Analysis of Weak Lines in APOGEE Spectra \citep[BAWLAS;][]{hayes22} are included; 
red clump bulge stars observed with UVES/VLT by
\citet{hill11}, and \citet{siqueira-mello16}. Other literature data include
bulge microlensed stars also observed with UVES/VLT by \citet{bensby17};
\citet{ryde16} observed stars in the inner 2$^{\circ}$ with the
CRyogenic high-resolution InfraRed Echelle Spectrograph $-$
CRIRES/VLT spectrograph in the K band, at a resolution of R $\sim$ 50,000. \citet{nandakumar24} observed a sample of bulge stars 1$^{\circ}$ north
 of the Galactic centre, using the
Immersion GRating INfrared Spectrograph IGRINS/Gemini spectrograph
at a resolution of R  $\sim$ 45,000 in the H and K bands.
We also include \citet{cunhasmith06} results, from the H band observed
with the Phoenix/Gemini spectrograph,
pointing out that they found high-Na abundances in two metal-rich
bulge stars.

From these two figures, we can conclude that: a) there seems to be a real scatter in the [Mg/Fe] ratios in the more metal-rich regime, with some stars showing enhanced ratios of [Mg/Fe]  at the highest metallicities, compared to the predictions of all models shown in the figure; b) there are large enhancements in [Na/Fe] for above-solar metallicities; and c) the [Al/Fe] ratios of the present work are now compatible with the other samples recently studied with high-resolution data.
The chemical evolution models shown in these figures indicate that:
a) with the new data, and given the errors for [Fe/H] and [Mg/H], the models can in general explain the observed [Mg/Fe] trend with [Fe/H] over the whole metallicity range, except for a number of stars with enhanced [Mg/Fe] at suprasolar metallicities, which are located above the upper envelope of the data distribution. The standard model lies in the middle or lower part of this region, but considering the increase by a factor of 10 in the Mg ejecta of SN Ia, the models  cover  most of the envelope. At [Fe/H] $\geq$ 0, there is a large [Mg/Fe] scatter, with stars presenting both very high and very low [Mg/Fe];
b) the models reproduce the observed [Na/Fe] versus [Fe/H] relation of metal poor stars, for which the standard model predicts a downward behaviour of this abundance ratio beyond [Fe/H] $\simeq$ $-$0.7 dex, but fail completely to explain the very large [Na/Fe] ratio observed in many metal-rich stars. For the Na abundances, increasing the Na ejected by SN Ia has a negligible impact on the predictions of the model. The present sample shows some stars with higher [Na/Fe] relative to \citet{bensby13,bensby17} and \citet{ernandes26}, recalling the Bensby's stars are 
mostly dwarfs, and the Na abundances by
\citet{ernandes26} were derived in the H band;
c) the models reproduce the [Al/Fe] results quite well,  now that our values are systematically lower than in \citet{lecureur07}. The scatter in the data seems to be well accounted for by the different models, although less so for the lower metallicities.

\begin{figure}
\includegraphics[angle=0,width=9cm]{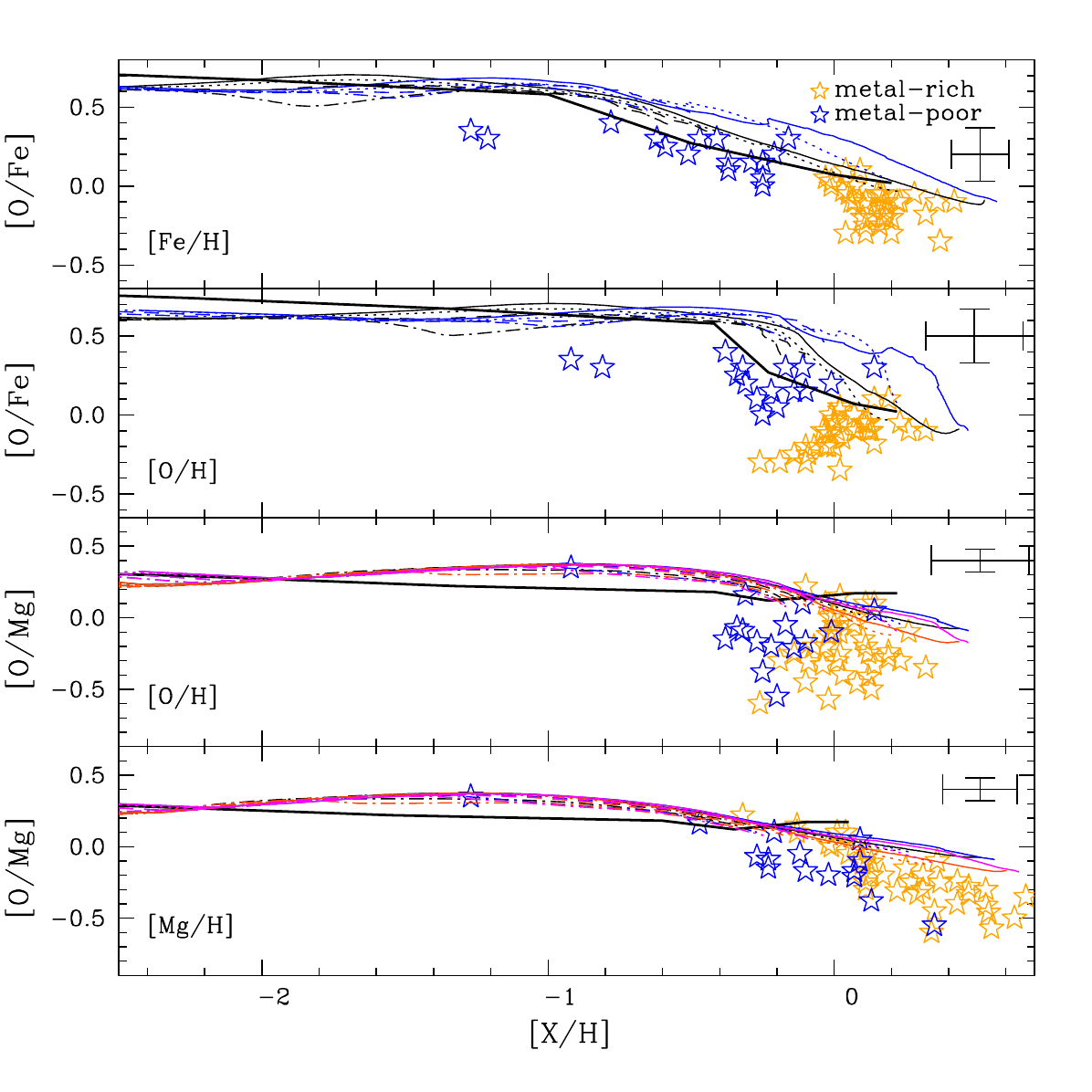}
\caption{\textit{Top}: [O/Fe] vs both [Fe/H] and [O/H]. \textit{Bottom}:  [O/Mg] ratio vs
both [O/H] and [Mg/H]. Chemical evolution models are the same as in Fig. \ref{jonsson}.
Thick black lines are the predictions of the K20 models.
The open orange symbols represent metal-rich stars and the open blue symbols metal-poor stars.
The error bars are the mean values from Table \ref{errors}.}
\label{plotOvsO}
\end{figure}

Figure \ref{plotOvsO} is intended to provide further insights into the nature of O and Mg enhancements.
The figure shows the [O/Fe] and [O/Mg] as functions of [Fe/H], [O/H], and [Mg/H].
To facilitate a comparison with other chemical evolution models, the K20 model of \citet{kobayashi20} is also exhibited.
In the [O/Fe]  versus [Fe/H] plot, the stars are clearly segregated in two groups: metal-rich stars with [Fe/H] $>-0.1$
and metal-poor stars  with [Fe/H] $<-0.1$.
The [O/Fe]  versus [Fe/H] relation shows the expected behaviour; namely, [O/Fe] decreases with increasing metallicity, as more and more type Ia supernovae have time to contribute to the chemical enrichment.
The model K20 displays the same behaviour as our model, with the difference that [O/Fe] falls more rapidly at intermediate metallicities than our chemodynamical model,
fitting the data in [O/Fe] well, but not the data in [O/Mg].
 The difference is due to a shorter timescale of bulge formation, so that the enrichment by SNe Ia occurs when there is still a large contribution from SNe II.
At [Fe/H] $<-1$, corresponding, for the model K20, to an early phase  of the disk formation, both models show a plateau in [O/Fe] at nearly the same level,
which is higher than the measured abundance ratios for the only two sample stars in this metallicity range.
For the [O/Fe]  versus [O/H] plot, the separator between the two metallically groups is now a diagonal, and the effect of the supernova appearance is even more abrupt for the K20 model in the solar neighborhood, but remains mild for the chemodynamical model.
The separation between metal-rich and metal-poor objects is less clear when plotting [O/Mg] as a function of [O/H], 
with the [O/Mg] ratio showing a large scatter, departing from solar, with many stars having [O/Mg] below solar in most of the metallicity range studied here,
while the abundance ratios predicted by the models are above or close to solar.
Interestingly, when plotting the same ratio as a function of [Mg/H], one sees that now the trend is better defined and that the metallicity range extends to larger values, with many stars very rich in Mg. This suggests that there is some mechanism that produces more Mg than O.

\begin{figure}
\includegraphics[angle=0,width=9cm]{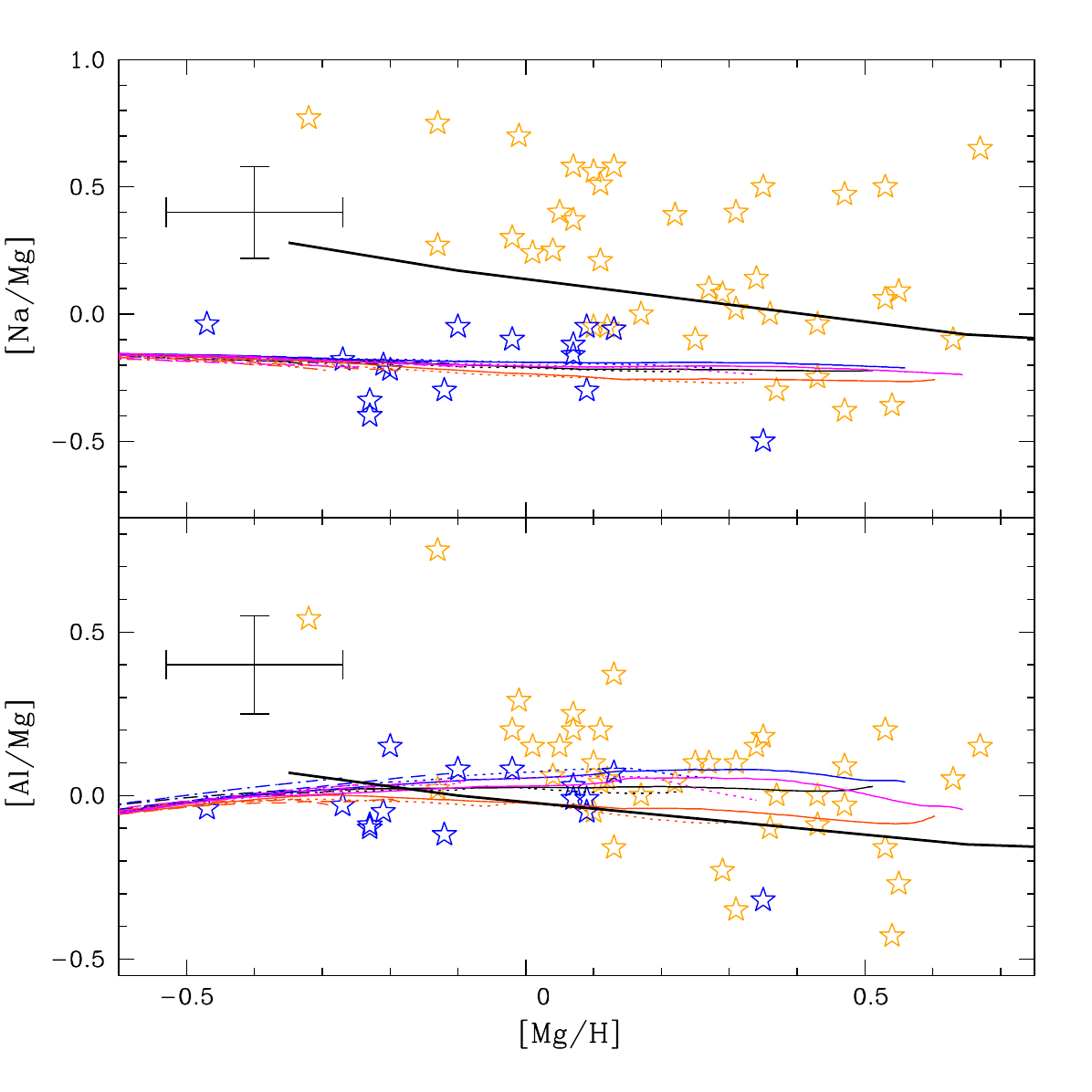}
\caption{[Na/Mg], and [Al/Mg] vs [Mg/H]  for the present sample. The symbols are the same as in Fig. \ref{plotOvsO}.
The chemical evolution models are the same as in previous figures.}
\label{plotnaalvsmg}
\end{figure}

Finally, we investigate the [Na/Mg] and [Al/Mg] versus [Mg/H] in Fig. \ref{plotnaalvsmg}. Here, our analysis leads  to the discovery of two stars with very large values of [Al/Mg], as well as many stars with large [Na/Mg] ratios, 
which are not predicted by the chemical evolution models.
The scatter in these plots seems real and is rather large. Note that stars with large [Na/Mg] now cover the entire [Mg/H] range, unlike in \citet{lecureur07} in which there was a tendency to have the largest [Na/Mg] towards the more Mg-rich stars.

\section{Discussion}\label{discussion}

Before interpreting the discrepancies highlighted in the previous section, it is useful to briefly recall the main nucleosynthetic channels responsible for the production of Na, Mg, and Al. The details of the stellar yields adopted in the models were described in the previous section, and here we focus  on the dominant physical processes and their expected dependencies on stellar mass and metallicity. The aim is to assess whether the observed abundance patterns in bulge giant stars can be understood within the context of current nucleosynthesis prescriptions or whether they point to missing ingredients in the models. Another possibility is that of the sample contamination by different stellar populations, but this will be analysed in Paper II.

The alpha-elements Mg and O are predominantly produced in massive stars, while Na and Al have contributions from both massive stars and AGB stars, 
with yields that are sensitive to metallicity.
According to WW95, Na is mainly synthesized during hydrostatic carbon burning, and its final abundance is also sensitive to the neutron excess,
whereas magnesium and aluminum are products of hydrostatic carbon and neon burning. In low- and intermediate-mass stars, Na can be produced
in the NeNa cycle \citep{denissenkov98}.
In terms of nucleosynthesis, the cause of larger Na/O production at a higher metallicity in massive stars is the variation in the neutron excess, $\eta$.
 A higher neutron excess produces larger amounts of the
odd-Z elements Na and Al. The neutron excess at carbon ignition is proportional to metallicity, Z, because during hydrogen and helium burning CNO is progressively transformed into  $^{14}$N, and further into $^{22}$Ne
through the reactions
$^{14}$N($\alpha$,$\gamma$)$^{18}$F(e$^+$$\nu$)$^{18}$O($\alpha$,$\gamma$)$^{22}$Ne. The decay of $^{18}$F produces a neutron excess where there was none, and that excess, $\eta$ = 0.0015 Z, persists and affects subsequent nucleosynthesis (Woosley, private communication).
In summary, in massive stars, a higher neutron excess favors the production of odd-Z elements, providing a natural explanation for the increase of Na/O and Al/O ratios; however, in this case, Mg is less directly affected. Note that the Na excess in metal-rich stars has also been reported for thin 
and thick disk stars 
\citep[e.g.][]{smiljanic16, bensby03,owusu26}.

Within this framework the abundance patterns shown in previous figures present some challenges. The high [Mg/Fe] values at super-solar metallicities, together with the large star-to-star scatter, are not reproduced by standard yields of core collapse supernovae. The behaviour of the [O/Mg] ratio provides an additional constraint: since O and Mg are expected to be co-produced in massive stars, the observed sub-solar [O/Mg] values and their large scatter suggest a relative overproduction of Mg with respect to O. Moreover, the chemical evolution models provide   a good fit to the [Al/Fe] ratios measured in this work.

\subsection{Type Ia supernovae}

Type Ia supernovae yields provide another alternative.
\citet{chiappini05} used Galactic disk chemical evolution models to test different SN Ia yield prescriptions and found that none of the models available at the time, including both 1D and 3D explosion models, could reproduce the high Mg abundances observed in metal-rich thin disk stars.
She suggested that SN Ia yields of Mg could be at least a factor of 10 higher than in classical models such as the 1D deflagration model W7 of \citet{iwamoto99} (our standard model adopts their W7 and W70 models).
Soon after, \citet{stehle05} reported measurements (using the technique of abundance tomography, which reconstructs the stratified chemical profiles of the ejecta from time-series spectra) of the chemical profiles of the type Ia SN2002bo. Table 3 in that study shows that the total ejected Mg is roughly one order of magnitude higher than in the W7 model.

Very interestingly, the same supernova also exhibits a similar enhancement in Na abundances with respect to W7 models. From the ratios of the masses ejected by SN 2002bo, 0.080, 0.001, and 0.880 $\ M_{\odot}$ of Mg, Na, and Fe, respectively, its ejecta have [Mg/Fe]= $-$0.76 and [Na/Fe]= $-$1.32. For comparison, the W7 model has [Mg/Fe]= $-$1.67 and [Na/Fe]= $-$2.45 (and the W70 model, [Mg/Fe]= $-$1.42 and [Na/Fe]= $-$3.11).
No Al was detected, as expected from the low abundances predicted by the classical SN Ia models (the models W7 and W70 show [Al/Fe]= $-$1.48 and $-$2.44, respectively).
It is worth noting that the 3D deflagration model b30\_3d\_768 from \citet{travaglio04}
that appears in Table 3 of \citet{stehle05} also shows a high Na abundance, [Na/Fe]= $-$1.16, even higher than that observed for SN 2002bo.
Another interesting point is that the mass of stable Ni isotopes derived for SN 2002bo (2.42$\times 10^{-2}$\ M$_{\odot}$) is much smaller than that predicted by the model W7 (1.26$\times 10^{-1}$\ M$_{\odot}$). The overestimated yields of Ni for the W7 models have already been taken into account by the models in \citet{barbuy24}.
In that paper, the W7 Ni yield has been divided by two because, as \citet{iwamoto99} pointed out,  this model overproduces
$^{58}$Ni, the main Ni isotope.

In view of the possibility that Mg, as well as Na yields, have been underestimated by the
\citet{iwamoto99} models, we ran the NaMg-enhanced models, in which both Mg and Na yields of W7 were multiplied by
a factor of 10. As we can see from Fig. \ref{plotmgnaalvsfe}, 
there is some improvement for the low-SFR-model for Mg, but the effect on the high-SFR-model for Mg and on both models for Na
is very small.

With respect to Mg enrichment, from the point of view of nucleosynthesis prescriptions, it is reasonable to
increase the Mg yields from \citet{iwamoto99}. The solar metallicity model W7, responsible for the chemical enrichment
in the metal-rich regime, shows the most severe depletion of Mg production.
Counterintuitively, the zero-metallicity model, model W70, exhibits both more iron (0.775 M$_{\odot}$), and more Mg
(0.0158 M$_{\odot}$) than the model 7 (0.749 M$_{\odot}$) and (0.00857 M$_{\odot}$), respectively, with the net
result that the W70 produces proportionally more Mg than Fe, implying [Mg/Fe]= $-$1.42,
and $-$1.67 for W70 and W7, respectively.
The higher ejected Fe mass for model W70 is due to the larger production of $^{56}$Ni.
The initial zero metallicity of W70 corresponds to Y$_e = 0.5$
(where Y$_e$ is the number of electrons per baryon), leading to the material in the white dwarf to
undergo complete Si burning, leaving $^{56}$Ni as the dominant nucleus, without competition by more neutron-rich species.
With respect to the higher Mg production in W70, the symmetric matter implies also a symmetric $\alpha$-burning sequence,
starting from $^{12}$C, halting at $^{24}$Mg, without being further converted into heavier neutron-rich nuclides.
On the other hand, the lack of neutron excess in W70, leads to an underproduction of odd-Z elements, such as Na and Al;
W70 has [Na/Fe]= $-$3.11, and [Al/Fe]= $-$2.49, whereas W7 gives [Na/Fe]= $-$2.45 and [Al/Fe]= $-$1.53.

Still, regarding Mg yields from SNe Ia, in recent years, there have been models with considerably higher Mg production rates than those from \citet{iwamoto99}.
 The off-centre delayed detonation (O-DOT) model from \citet{maeda10} has
[Mg/Fe]= $-$1.21 and the 2D detonating failed deflagration model  Y12 from \citet{plewa07} [Mg/Fe]= $-$0.75,
but the most promising modern models are those involving sub-Chandrasekhar-mass white dwarfs:
the two-explosion sub-Chandrasekhar model from \citet{pakmor22}, with  [Mg/Fe]= $-$0.85,
and a very comprehensive grid of sub-Chandrasekhar models presented by \citet{leungnomoto20} for
white dwarf masses in the range of 0.9$-$1.2 M$_{\odot}$, where the 0.9 M$_{\odot}$ models exhibit suprasolar [Mg/Fe] values.
For instance, the model 090-050-2-B50 has [$^{24}$Mg/$^{56}$Fe]= 0.55  (but increasing the white dwarf mass to 1 M$_{\odot}$, [$^{24}$Mg/$^{56}$Fe] drops to $-$0.98).
  The explosion of sub-Chandrasekhar models has thinner burning layers in its structure, producing a higher relative
  ratio of intermediate-mass elements such as Mg, Si, and S over Fe \citep{leungnomoto23}.
Masses from 0.05 to 0.15 M$_{\odot}$ of Mg can be produced in the sub-Chandrasekhar SN Ia scenario.

On the other hand, considering the Na enhancement, all SN Ia models found in the literature have strongly subsolar [Na/Fe].
A recent GALAH (GALactic Archaeology with HERMES) study of disk stars \citep{owusu26} 
investigated the role of SNe Ia with respect to the high [Na/Fe] found at high metallicities and concluded that even
the most favourable sub-Chandrasekhar-mass SN Ia has a negligible contribution of [$^{23}$Na/$^{56}$Fe]= $-$1.78.
Even the extreme model 090-050-2-B50 from \citet{leungnomoto20} has [$^{23}$Na/$^{56}$Fe]= $-$0.95.

Finally, when interpreting Figs. \ref{jonsson} and \ref{plotmgnaalvsfe}, it is important to recall that the contribution
from SNe Ia is to lower the [X/Fe] models with increasing metallicity, because they dilute intermediate metal elements relative to Fe,
since their main contribution is Fe. In other words, the ejecta from SNe Ia are almost always subsolar; therefore, they would never make [X/Fe] suprasolar.

\subsection{Massive rotating stars}

The high-resolution data show that [Mg/Fe] remains elevated and scattered at super-solar metallicities (Figs. \ref{jonsson} and \ref{plotmgnaalvsfe} ), while [Na/Fe] increases with increasing metallicity, contrary to the trend predicted by standard chemical evolution models. As we noted before, the [O/Mg] ratios in Fig. \ref{plotOvsO} reveal a relative overproduction of Mg with respect to O. A possible way to decrease O/Mg with increasing metallicity can be related to the nucleosynthesis of massive stars. Metal-dependent massive-star yields arise primarily from metallicity-dependent stellar structure and mass loss, which preferentially removes oxygen-rich layers and alters the relative contributions of hydrostatic burning zones.
Consequently, oxygen yields are more strongly suppressed than magnesium yields, producing a declining O/Mg trend. In massive stars rotation leads to a higher neutron excess, favouring the production of odd-Z elements, providing a natural explanation for the increase in Na/O
and Al/O ratios. In addition, while massive fast-rotating stars at low metallicity can produce non-negligible heavy s-process elements, including Ce, this channel becomes inefficient at solar and super-solar metallicity due to the increased seed abundance and reduced neutron-to-seed ratios, limiting the production to the weak s-process regime. 
This expected behaviour is confirmed in \citet{ernandes26}, in which it is shown that [Ce/Fe] decreases with increasing metallicity.

\subsection{AGB and super-AGB contribution}

The Na excess can be caused by yields from intermediate-mass AGB stars combined
with metallicity-dependent yields from massive stars.
In particular, AGBs produce nitrogen and Na, increasing as a function of metallicity according
to \citet{ventura13}, \citet{karakaslattanzio14},  and \citet{cristallo15}.

In a recent study \citep{cinquegrana22}, yields from metal-rich
AGBs (0.04 $\leq$ Z $\leq$ 0.10) were presented, where a trend of higher Na production with increasing metallicity is confirmed.
Furthermore, even higher yields for suprasolar metallicities are predicted. Also super-AGBs produce higher Na. As an example, the super AGB model (8 M$_{\odot}$) of \citep{cinquegrana22} of Z=0.04 produces 
5.6 $\times$ 10$^{-4}$ M$_{\odot}$ of Na. A typical mass AGB 
(5 M$_{\odot}$) for the same metallicity also releases a large
amount of Na (2.0 $\times$ 10$^{-4}$ M$_{\odot}$).

The timescales for bulge formation are long enough in our chemodynamical model
($\sim$ 0.3 to 1 Gyr) for the emergence of AGBs,
which appear at 30$-$40 Myr, in the case of super-AGBs, which have an initial
mass in the range of 6$-$10 M$_{\odot}$, and in 100$-$300 Myr for a typical AGB, which has a mass of 3$-$5 M$_{\odot}$.
There is also time for contributions of even lower mass
(2$-$3 M$_{\odot}$), which emerge in 0.3$-$1 Gyr after star formation.
In conclusion, AGBs could play a key role in the chemical evolution of Na in the Galactic bulge.
It should be noted also that there is a crucial difference between the
chemical evolution of Na (and other intermediate-mass elements)
between AGBs and SNe Ia: the latter dilute these elements relative to Fe, since their main contribution is Fe, whereas AGBs only
contribute with the intermediate-mass elements, and not Fe, causing an increase in their abundances relative to Fe.

\subsection{Globular cluster second-generation stars}

\begin{figure}
\includegraphics[angle=0,width=9cm]{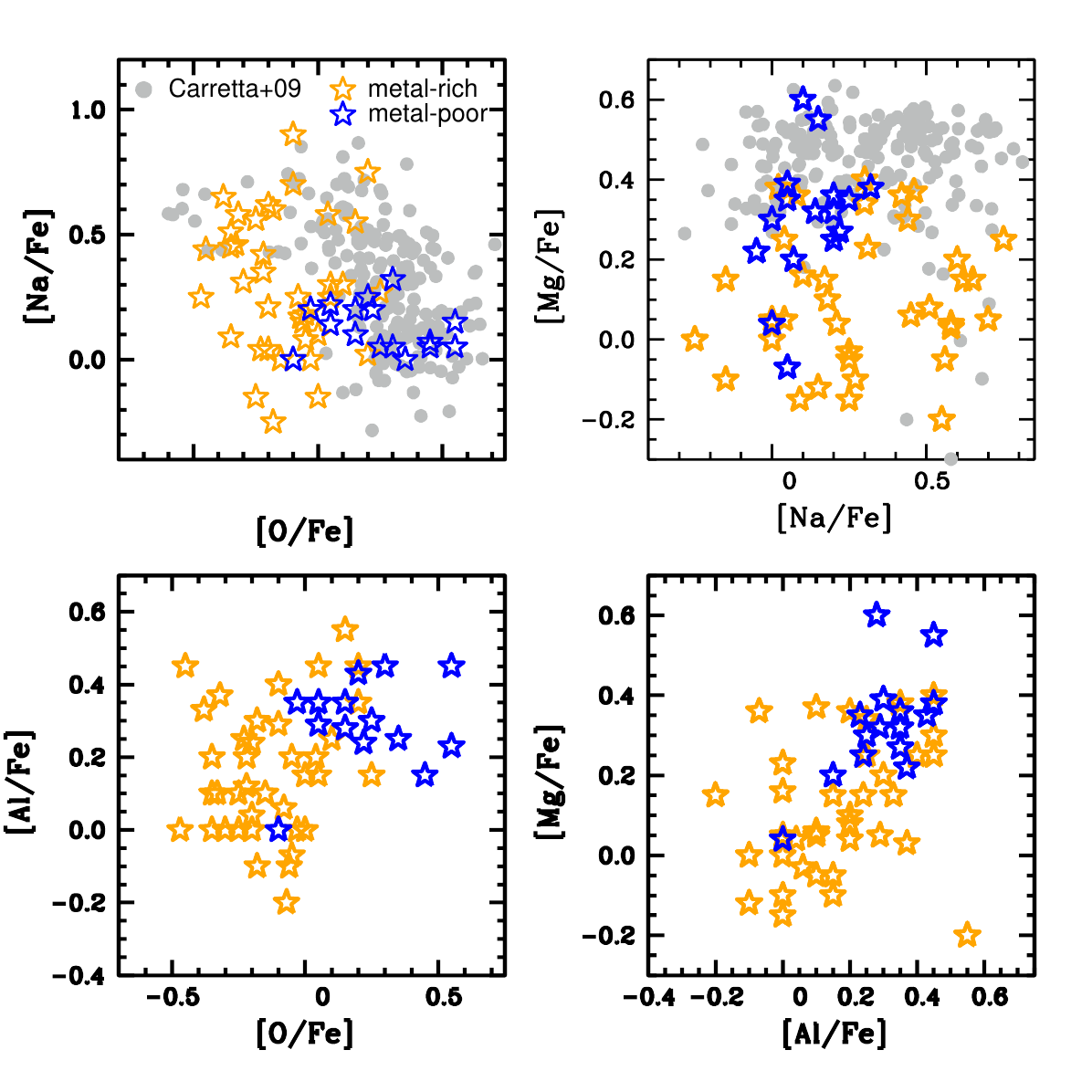}
\caption{[Na/Fe] vs [O/Fe] (\textit{top left}), [Mg/Fe] vs [Na/Fe] (\textit{top right}) for the present
sample compared with GC abundances from \citet{carretta09a}, showing a possible anti-correlation; 
[Al/Fe] vs [O/Fe] (\textit{bottom left}) and [Mg/Fe] vs [Al/Fe] (\textit{bottom right}) for the present sample,
showing a possible correlation.
The open orange stars represent metal-rich stars, the open blue stars metal-poor stars, and the filled grey circles
GC stars from \citet{carretta09a}.}
\label{alvio}
\end{figure}

From chemical evolution it is expected that [O/Fe]$<$0 for [Fe/H]$>$0.
We also know that in GCs, second-generation stars often
have low O and high Na and Al, while Mg is only mildly deficient \citep{carretta10}. This results from
proton-capture reactions of the CNO, NeNa, and MgAl
chains during H-burning at high temperature \citep{renzini15}.
In this work we would like to understand why the metal-rich stars show
enhanced Na abundances, and this could be explained by both circumstances.
In Fig. \ref{alvio} we plot [Na/Fe] versus [O/Fe], and [Mg/Fe] versus [Na/Fe],
compared with the data for 203 stars in 15 GCs, from
\citet{carretta09a,carretta09b}, to check if our sample behaves similarly
to the Na-O and Mg-Na anti-correlation as seen in the globulars.
It appears that for the metal-rich stars an anti-correlation Na-O could be
present, since the behaviour resembles that of the globular
clusters, although with accentuated O-deficiency.
We also plot [Al/Fe] versus [O/Fe] and [Mg/Fe] versus [Al/Fe], which appear
to be correlated, and not anti-correlated, for the metal-rich stars, making this interpretation less clear.
In this hypothesis, i.e. if stars with the Na enhancement are all second-generation
stars from GCs, it remains to be understood why this appears only in
the metal-rich stars.

\section{Conclusions}\label{conclusions}

 We re-derived the abundances of Na, Mg, and Al in a sample of 55
 bulge red giants, previously studied by \citet{zoccali06,lecureur07},
 and \citet{hill11}. We also re-derived the abundances of C, N, and O, for consistency
 in the use of the code \textsc{Turbospectrum}.
 For this well-studied sample, we had previously derived abundances of
 C, N, O, Na, Mg, Al, Mn, Zn, Co, Cu, and heavy elements.
  
The main aim of this study is the verification of overabundances of Na, Mg, and Al in the metal-rich bulge stars, using the code \textsc{Turbospectrum}, and updated atomic and molecular line
lists. 
 
\begin{itemize}
    \item Some metal-rich stars have confirmed excesses of Na, Mg, and Al.
    \item $\text{[Na/Fe]}$ overabundance, in particular, is confirmed, even if it is shown to be dependent on the adopted stellar parameters (we show abundances for two sets of input stellar parameters).
     \item Sub-solar and scattered [O/Mg] ratios indicate a relative overproduction of Mg compared to O, which could point to both a large Mg production by prompt SN Ia and metallicity-dependent yields of  core collapse stellar yields (linked to neutron excesses).
      \item The revised [Al/Fe] ratios are systematically lower than the original ones, and well reproduced by chemical evolution models, suggesting that no major revision is needed for Al yields. 
      \item Globular cluster-like enrichment 
      is clear for two stars in particular, found to show strongly enhanced [Al/Fe] ratios. Additionally, a Na-O anti-correlation
      appears to be present for the metal-rich stars, reinforcing the possibility of Na-enhanced stars being second-generation
      stars that have escaped from GCs.
       \item The enhancement of  Mg, but not for Na, at the metal-rich end, could be explained by increased enrichment by SNe Ia.
    \item The possible mechanisms for Na-enrichment include metallicity-dependent yields from massive stars, AGB stars, and
    an unknown mechanism that produces second-generation stars in GCs.
\end{itemize}

\begin{acknowledgements}

We are grateful to Stan Woosley for very helpful discussions, as well as calculations in progress. We are also grateful
to the referee, Ulrike Heiter, for a detailed report, that greatly improved the manuscript.
R.P.N. acknowledges a PhD Fapesp fellowship no.  2025/28071-6.
C.C. acknowledges partial financial support from FAPESP.
M.C.J acknowledges a Inicia\c c\~ao Cient\'{\i}fica PIBIC/CNPq fellowship.
 B.B. and A.C.S.F. acknowledge partial financial support from the      brazilian agencies CAPES - Financial code 001, CNPq and FAPESP. 
M.Z. acknowledges support by ANID FONDECYT regular 1230731 and by ANID BASAL Center for Astrophysics and Associated Technologies (CATA) through grant FB210003
A.P.-V. acknowledges the DGAPA–PAPIIT grant IN112526.
Observations were collected at the European  Southern  Observatory,  Paranal,  Chile  (ESO programmes  71.B-0617A, 73.B0074A, and GTO 71.B-0196) 
\end{acknowledgements}

\bibliographystyle{aa} 
\bibliography{bibliogna}

\begin{appendix}

\section{Sample stars: Basic characteristics and resulting abundances}

Table \ref{atmos} reports the
list of sample stars, their atmospheric parameters and radial velocities  adopted from \citet{zoccali06}, \citet{lecureur07}, and \citet{hill11}, and resulting C, N, O, Na, Mg, and Al abundances.

    \begin{table*}
\begin{flushleft}
\scalefont{0.8}
\caption{Sample stars: Basic characteristics and resulting abundances.}            
\label{atmos}
\resizebox{\textwidth}{!}{
\begin{tabular}{l c c c c c c c c c c c c c c c}
\noalign{\smallskip}
\hline\hline
\noalign{\smallskip}
\noalign{\vskip 0.1cm}
Star & OGLE no. & Gaia DR3 & $\alpha$(J2000) & $\delta$(J2000) & T$_{\rm eff}$ & log~g & [Fe/H] & v$_{\rm t}$ & vr$_{\rm helio}$ & [C/Fe] & [N/Fe] & [O/Fe] & [Na/Fe] & [Mg/Fe] & [Al/Fe] \\
\noalign{\vskip 0.1cm}
 &  &  & (h:m:s) & ($^{\circ}$:':") & (K) & (cgs) &  & (km.s$^{-1}$0  & (km.s$^{-1}$) & &  &  &  &  &  \\
\noalign{\vskip 0.1cm}
\noalign{\hrule\vskip 0.1cm}
\noalign{\vskip 0.1cm}
\noalign{\hrule\vskip 0.1cm}
\noalign{\hrule\vskip 0.1cm}
\noalign{\vskip 0.1cm}
B6-b1 & 29280c3 & 4049053136672718208 & 18 09 50.480 & $-$31 40 51.61 & 4400 & 1.8 & 0.07 & 1.6 
&$-$92.66$\pm$0.18   
& $-$0.10  & +0.30  &+0.04 & +0.58 &  +0.04 & +0.20 \\
B6-b2 & 83500c6 & 4048860859576854272 & 18 10 33.980 & $-$31 49 09.15 & 4200 & 1.5 &$-$0.01 & 1.4 
& 12.065$\pm$0.23  & $-$0.20  & +0.25 &$-$0.10  & +0.70  & +0.05 & +0.29 \\
B6-b3 & 31220c2 & 4048866975610770432 & 18 10 19.060 & $-$31 40 28.19 & 4700 & 2.0 & 0.10 & 1.6 
&$-$150.16$\pm$0.24  & $-$0.25 & +0.30  &$-$0.25 & +0.56 & $-$0.05 & +0.10 \\
B6-b4 & 60208c7 & 4048836945198766976 & 18 10 07.770 & $-$31 52 41.36  & 4400 & 1.9 &$-$0.41 & 1.7 
& $-$24.63$\pm$0.07  & $-$0.15 & +0.25 &+0.45 & +0.07 & +0.20 & +0.15 \\
B6-b5 & 31090c2 & 4048866254056263040 & 18 10 37.380 & $-$31 40 29.14 & 4600 & 1.9 &$-$0.37 & 1.3 
& $-$8.62$\pm$0.11   & $-$0.10  & +0.40  &+0.20  & +0.25 & +0.35 & +0.43 \\
B6-b6 & 77743c7 & 4048839934496141568 & 18 09 49.100 & $-$31 50 07.66 & 4600 & 1.9 & 0.11 & 1.8 
& 39.66$\pm$0.18   & $-$0.15 & +0.40  &$-$0.18 & +0.60 & +0.20 & +0.30 \\
B6-b8 & 108051c7 & 4048841308885921280 & 18 09 55.950 & $-$31 45 46.33 & 4100 & 1.6 & 0.03 & 1.3 
&$-$114.358$\pm$0.20 & +0.10  & $-$0.10  &+0.05 &  +0.25 & $-$0.05 & +0.15 \\
B6-f1 & 23017c3 & 4048865601221169280 & 18 10 04.460 & $-$31 41 45.31 & 4200 & 1.6 &$-$0.01 & 1.5 
& 34.34$\pm$0.20   & $-$0.10  & +0.28 &$-$0.06 &  +0.15 & $-$0.12 & $-$0.10 \\
B6-f2 & 90337c7 & 4048863608355972608 & 18 10 11.510 & $-$31 48 19.28 & 4700 & 1.7 &$-$0.51 & 1.5 
& $-$102.55$\pm$0.10 & +0.00  & $-$0.10  &$-$0.10  & +0.00 & +0.04  &  +0.00 \\
B6-f3 & 21259c2 & 4048865395070315904 & 18 10 17.720 & $-$31 41 55.20 & 4800 & 1.9 &$-$0.29 & 1.3 
& 86.06$\pm$0.10   & $-$0.10  & +0.00  &$-$0.03 & +0.20 &+0.36   &  +0.35 \\
B6-f5 & 33058c2 & 4048866425862504704 & 18 10 41.510 & $-$31 40 11.88 & 4500 & 1.8 &$-$0.37 & 1.4 
& 17.85$\pm$0.12   & +0.00  & +0.00  &+0.05 & +0.22 & +0.27  &  +0.35 \\
B6-f7 & 100047c6 & 4048861718570450048 & 18 10 52.300 & $-$31 46 42.18 & 4300 & 1.7 &$-$0.42 & 1.6 
& $-$14.44$\pm$0.11  & +0.00  & +0.20  & ---  & $-$0.05 & +0.22  & +0.37 \\
B6-f8 & 11653c3 & 4048865085825004928 & 18 09 56.840 & $-$31 43 22.56 & 4900 & 1.8 & 0.04 & 1.6 
& 54.56$\pm$0.17   & $-$0.25 & +0.35 &$-$0.45 & +0.44 & +0.30  &  +0.45 \\
BW-b2 & 214192 & 4050185874168661760 & 18 04 23.950 & $-$30 05 57.80  & 4300 & 1.9 & 0.22 & 1.5 
& $-$27.59$\pm$0.18  & +0.00  & +0.25 & +0.0 & $-$0.15 & +0.15 &  +0.15 \\
BW-b4 & 545277 & 4050186355212762624 & 18 04 05.340 & $-$30 05 52.50  & 4300 & 1.4 & 0.07 & 1.4 
& 77.33$\pm$0.35   & +0.15 & +0.30  &+0.15 & +0.55 & $-$0.20 & +0.55 \\
BW-b5 & 82760 & 4050205119984310144 & 18 04 13.270 & $-$29 58 17.80  & 4000 & 1.6 & 0.17 & 1.2 
& 60.65$\pm$0.26   & +0.20  & +0.15 &+0.25 & +0.27 & $-$0.10 & +0.15 \\
BW-b6 & 392931 & 4050185152723763712 & 18 03 51.840 & $-$30 06 27.90  & 4200 & 1.7 &$-$0.25 & 1.3 
& 132.20$\pm$0.11  & +0.00  & +0.65 &+0.15 & +0.20 & +0.32 & +0.35 \\
BW-b7 & 554694 & 4050187214332365056 & 18 04 04.570 & $-$30 02 39.60  & 4200 & 1.4 & 0.10 & 1.2 
& $-$219.23$\pm$0.38 & $-$0.05 & $-$0.15 &$-$0.22 & +0.35 & $-$0.42 & +0.12 \\
BW-f1 & 433669 & 4050203741234555392 & 18 03 37.140 & $-$29 54 22.30  & 4400 & 1.8 & 0.32 & 1.6 
& 194.64$\pm$0.43 & $-$0.10  & +0.10  & $-$0.20  & +0.62 & +0.15 & +0.24 \\
BW-f4 & 537070 & 4050183881416931072 & 18 04 01.400 & $-$30 10 20.70  & 4800 & 1.9 &$-$1.21 & 1.7 
& $-$152.30$\pm$1.59 & +0.10  & +0.30  & +0.30 & +0.05  & --- & --- \\
BW-f5 & 240260 & 4050195391975568640 & 18 04 39.620 & $-$29 55 19.80  & 4800 & 1.9 &$-$0.59 & 1.3 
& $-$13.94$\pm$0.94 & +0.10  & +0.40  &+0.05 & +0.14 & +0.32 & +0.29 \\
BW-f6 & 392918 & 4050196456967930240 & 18 03 36.890 & $-$30 07 04.30  & 4100 & 1.7 &$-$0.21 & 1.5 
& 174.38$\pm$0.59  & +0.00  & +0.20  &+0.35 & +0.00 & +0.30 & +0.25 \\
BW-f7 & 357480 & 4050189035264998016 & 18 04 43.920 & $-$30 03 15.20  & 4400 & 1.9 & 0.11 & 1.7 
& $-$147.17$\pm$0.35 & $-$0.25 & +0.60  & $-$0.20 & +0.21 & +0.04 & +0.04 \\
BW-f8 & 244598 & 4050198278034975872 & 18 03 30.490 & $-$30 01 44.80  & 5000 & 2.2 &$-$1.27 & 1.8 
& $-$32.98$\pm$2.57  & +0.00  & +0.20  & +0.45  & +0.05  & -0.07 & --- \\
BL-1 & 1458c3 & 6734687910490382720 & 18 34 58.643& $-$34 33 15.241     & 4500 & 2.1 &$-$0.16 & 1.5 
& 77.89$\pm$0.07   & +0.00  & +0.50  & +0.22 & +0.20 & +0.25 & +0.24 \\
BL-3 & 1859c2 & 6734735052049421952 & 18 35 27.640 & $-$34 31 59.353    & 4500 & 2.3 &$-$0.03 & 1.4 
& 21.85$\pm$0.14   & +0.05 & +0.00  & +0.00  & +0.10 & +0.16 & +0.00 \\
BL-4 & 3328c6 & 6734635610644904832 & 18 35 21.240& $-$34 44 48.217     & 4700 & 2.0 & 0.13 & 1.5 
& 89.20$\pm$0.13   & $-$0.15 & +0.20  & $-$0.32 & +0.58 & +0.02 & +0.37 \\
BL-5 & 1932c2 & 6734735773603789568 & 18 36 01.148& $-$34 31 47.913     & 4500 & 2.1 & 0.16 & 1.6 
& 29.16$\pm$0.19   & $-$0.25 & +0.20  & $-$0.35 & +0.09 & $-$0.15 & +0.00 \\
BL-7 & 6336c7 & 6734730447844247040 & 18 35 57.392& $-$34 38 04.621     & 4700 & 2.4 &$-$0.47 & 1.4 
& 79.47$\pm$0.07   & +0.15 & +0.50  & +0.55 & +0.05 & +0.35 & +0.23 \\
B3-b1 & 132160C4 & 4064893495841461888 & 18 08 15.840 & $-$25 42 09.83  & 4300 & 1.7 &$-$0.78 & 1.5 
&$-$151.44$\pm$0.10  &  ---  &  ---  & +0.55 & +0.15 & +0.55 & +0.45 \\
B3-b2 & 262018C7 & 4064836493015520384 & 18 09 14.062 & $-$25 56 47.35  & 4500 & 2.0 & 0.18 & 1.5 
& $-$19.70$\pm$0.08  & $-$0.10  & +0.30  & $-$0.05 & +0.08 & +0.36 & $-$0.07 \\
B3-b3 & 90065C3 & 4064890055484375168 & 18 08 46.405 & $-$25 42 44.40   & 4400 & 2.0 & 0.18 & 1.5 
& $-$15.22$\pm$0.29  & $-$0.10  & +0.05 & $-$0.33 & +0.46 & +0.37 & +0.10 \\
B3-b4 & 215681C6 & 4064837798687212928 & 18 08 44.472 & $-$25 57 56.85  & 4500 & 2.1 & 0.17 & 1.7 
& 51.02$\pm$0.08   & +0.00  & +0.17 & $-$0.22 & +0.42 & +0.36 & +0.20 \\
B3-b5 & 286252C7 & 4064888474938466048 & 18 09 00.527 & $-$25 48 06.78  & 4600 & 2.0 & 0.11 & 1.5 
& $-$78.80$\pm$0.24  & $-$0.25 & +0.20  & $-$0.35 & +0.51 & +0.08 & +0.20 \\
B3-b7 & 282804C7 & 4064841337739525248 & 18 09 16.540 & $-$25 49 26.08  & 4400 & 1.9 & 0.20 & 1.3 
& 132.23$\pm$0.21  & $-$0.35 & +0.25 & $-$0.47 & +0.25 & $-$0.15 & +0.00 \\
B3-b8 & 240083C6 & 4064887757763678208 & 18 08 24.602 & $-$25 48 44.39  & 4400 & 1.8 &$-$0.62 & 1.4 
& $-$37.05$\pm$0.13  & $-$0.15 & +0.15 & +0.25 & +0.05 & +0.39 & +0.30 \\
B3-f1 & 129499C4 & 4064893388378926080 & 18 08 16.176 & $-$25 43 19.18  & 4500 & 1.9 & 0.04 & 1.6 
& 3.47$\pm$0.22    & $-$0.15 & +0.15 & $-$0.08 & +0.25 & $-$0.01 & +0.06 \\
B3-f2 & 259922C7 & 4064836282491581184 & 18 09 15.609 & $-$25 57 32.75  & 4600 & 1.9 &$-$0.25 & 1.8 
& $-$22.55$\pm$0.11  & +0.25 & +0.40  & +0.30  & +0.32 & +0.38 & +0.45 \\
B3-f3 & 95424C3 & 4064898924682446976 & 18 08 49.628 & $-$25 40 36.93   & 4400 & 1.9 & 0.06 & 1.7 
& $-$45.03$\pm$0.27  & $-$0.15 & +0.05 & $-$0.30  & +0.31 & +0.23 & +0.00 \\
B3-f4 & 208959C6 & 4064837386369122176 & 18 08 44.293 & $-$26 00 25.05  & 4400 & 2.1 & 0.09 & 1.5 
& $-$107.79$\pm$0.09 & +0.10  & +0.10  & +0.20  & +0.02 & +0.38 & +0.35 \\
B3-f5 & 49289C2 & 4064895042030572928 & 18 09 18.404 & $-$25 43 37.41   & 4200 & 2.0 & 0.16 & 1.8 
& $-$60.64$\pm$0.26  & +0.00 & +0.20  & $-$0.07 & +0.17 & +0.15 & $-$0.20 \\
B3-f7 & 279577C7 & 4064840302585112832 & 18 09 23.694 & $-$25 50 38.19  & 4800 & 2.1 & 0.16 & 1.7 
& $-$35.13$\pm$0.16  & $-$0.25 & +0.35 & $-$0.35 & +0.45 & +0.06 & +0.10 \\
B3-f8 & 193190C5 & 4064887276728723200 & 18 08 12.632 & $-$25 50 04.45  & 4800 & 1.9 & 0.20 & 1.5 
& $-$14.90$\pm$0.27  & $-$0.15 & +0.25 & $-$0.38 & +0.65 & +0.15 & +0.33 \\
\hline
\hline
BWc-1  & 393125 & 4050196938012154240 & 18 03 50.445  & $-$30 05 31.993 & 4476 & 2.1 & 0.09 & 1.5 
& 111.8$\pm$0.91   & +0.00  & +0.30  & +0.10  & +0.30 & +0.34 & +0.25 \\
BWc-2  & 545749 & 4050186595730916224 & 18 03 56.824  & $-$30 05 37.390 & 4558 & 2.2 & 0.18 & 1.2 
& 62.6$\pm$0.93    & $-$0.20  & +0.08 &$-$0.23  & +0.04 & +0.25 & +0.25 \\
BWc-3  & 564840 & 4050198866509717120 & 18 03 54.730  & $-$30 01 06.096 & 4513 & 2.1 & 0.28 & 1.3 
& 237.6$\pm$0.70   & $-$0.15 & +0.70  & +0.20  & +0.75 & +0.25 & +0.45 \\
BWc-4  & 564857 & 4050198866509710592 & 18 03 55.416  & $-$30 00 57.314 & 4866 & 2.2 & 0.05 & 1.3 
& 1.1$\pm$0.21     & $-$0.20  & +0.00 & $-$0.20  & +0.04 & +0.05 & +0.00 \\
BWc-5  & 575542 & 4050206386940903936 & 18 03 56.021  & $-$29 55 43.716 & 4535 & 2.1 & 0.42 & 1.5 
& 65.0$\pm$0.67    & $-$0.08 & +0.38 & $-$0.10  & +0.90 & +0.25 & +0.40 \\
BWc-6  & 575585 & 4050206386934265600 & 18 03 56.543  & $-$29 55 11.787 & 4769 & 2.2 &$-$0.25 & 1.3 
& 104.9$\pm$       & +0.00  & +0.05 & +0.15 & +0.10 & +0.60 & +0.28 \\
BWc-8  & 78255  & 4050200653215012992 & 18 03 12.494  & $-$30 03 59.111 & 4610 & 2.2 & 0.37 & 1.3 
& $-$4.2$\pm$0.97    & $-$0.30  & +0.15 & $-$0.25 & $-$0.15 & $-$0.10 & +0.00 \\
BWc-9  & 78271  & 4050197801356652032 & 18 03 16.683  & $-$30 03 51.406 & 4539 & 2.1 & 0.15 & 1.5 
& 47.8$\pm$0.97    & $-$0.10  & +0.15 & $-$0.05 & +0.18 & +0.10 & +0.20 \\
BWc-10 & 89589  & 4050201237280480128 & 18 03 18.914  & $-$30 01 09.983 & 4793 & 2.2 & 0.07 & 1.3 
& 188.0$\pm$1.24   & $-$0.15 & +0.12 & $-$0.15 & +0.00 & +0.05 & +0.10 \\
BWc-11 & 89735  & 4050201958885654912 & 18 03 04.749  & $-$29 59 35.301 & 4576 & 2.1 & 0.17 & 1.0 
& 98.0$\pm$1.17    & $-$0.20  & +0.07 & $-$0.03 & +0.00 & +0.00 & +0.00 \\
BWc-12 & 89832  & 4050201684007539712 & 18 03 20.102  & $-$29 58 25.785 & 4547 & 2.1 & 0.23 & 1.3 
& $-$47.6$\pm$0.84   & +0.00  & +0.10  & +0.05 & +0.30 & +0.40 & +0.45 \\
BWc-13 & 89848  & 4050202538650118656 & 18 03 04.612  & $-$29 58 14.080 & 4584 & 2.1 & 0.36 & 1.1 
& $-$201.1$\pm$1.01  & $-$0.15 & +0.00  & $-$0.18 & $-$0.25 & +0.00 & $-$0.10 \\
\noalign{\vskip 0.1cm}
\noalign{\hrule\vskip 0.1cm}
\noalign{\vskip 0.1cm}
\hline
\end{tabular}
}
\end{flushleft}
\end{table*}  

\section{Comparison with other work}

Table \ref{jonsson} reposts a 
comparison of stellar parameters, and Mg, Na, and Al abundances 
computed with stellar parameters from \citet{johnson14} and \citet{jonsson17} stellar parameters.

\begin{table*}
\scalefont{0.7}
\caption{Comparison of Mg, Na, and Al abundances with other stellar parameters. }   
\label{johnson} 
\centering                  
\begin{tabular}{c c c c c c | c c c c c c c} 
\hline\hline  
Star &   OGLE  & T$_{\rm eff}$ & log~g  & [Fe/H]  & v$_{\rm t}$  & T$_{\rm eff}$ & logg & [Fe/H]& v$_{\rm t}$& [Na/Fe] & [Mg/Fe] & [Al/Fe]   \\
     &     & K  & cgs  &   & km.s$^{-1}$  & K & cgs & & km.s$^{-1}$ &  &  &    \\
\hline
 &   &  \multicolumn{4}{c}{\citet{lecureur07}}  &   \multicolumn{4}{c}{\citet{johnson14}}  &  &  &   \\
\hline  
B3-b2 &  262018C7 & 4500 &  2.0 &   0.18  &  1.5 &  4700 &  2.75 &  0.08 & --- & +0.51 & +0.55  & +0.25   \\
B3-b4 &  215681C6 & 4500 &  2.1 &   0.17  &  1.7 &  4800 &  2.75 &  0.31 & --- & +0.65 & +0.20  & +0.46  \\
B3-b5 &  286252C7 & 4600 &  2.0 &   0.11  &  1.5 &  4700 &  3.10 &  0.43 & --- & +0.60 & +0.12  & +0.24   \\
B3-b7 &  282804C7 & 4400 &  1.9 &   0.20  &  1.3 &  4575 &  2.50 &  0.10 & --- & +0.80 & +0.26  & +0.42   \\
B3-b8 &  240083C6 & 4400 &  1.8 &   $-$0.62 &  1.4 &  4425 &  1.65 & $-$0.58 & --- & +0.12 & +0.23  & +0.31   \\
B3-f1 &  129499C4 & 4500 &  1.9 &   0.04  &  1.6 &  4900 &  2.75 &  0.13 & --- & +0.62 & +0.24  & +0.37   \\
B3-f8 &  193190C5 & 4800 &  1.9 &   0.20  &  1.5 &  4675 &  2.75 &  0.24 & --- & +0.84 & +0.20  & +0.42   \\
\hline
\hline 
&   &  \multicolumn{4}{c}{\citet{lecureur07}}  &   \multicolumn{4}{c}{\citet{jonsson17}}  &  &  &   \\
\hline
B3-b1& 132160C4 & 4300 & 1.7 & $-$0.78 & 1.5& 4414& 1.35& $-$0.89 &  1.41 & +0.33  & +0.70   & +0.66   \\
B3-b5& 286252C7 & 4600 & 2.0 & 0.11 & 1.5 & 4425& 2.70& 0.25 &  1.43 & +0.45  & +0.20  & +0.10    \\
B3-b7& 282804C7 & 4400 & 1.9 & 0.20 & 1.3 & 4303& 2.36& 0.08 &  1.58 & +0.41  & +0.21 & +0.19 \\
B3-b8& 240083C6 & 4400 & 1.8 & $-$0.62 & 1.4& 4287& 1.79& $-$0.67& 1.46 & +0.04 & +0.44  & +0.34  \\
B3-f1& 129499C4 & 4500 & 1.9 & 0.04 & 1.6 & 4485& 2.25& $-$0.15& 1.88 & +0.45 & +0.37 & +0.36 \\
B3-f2& 259922C7 & 4600 & 1.9 & $-$0.25 & 1.8& 4207& 1.64& $-$0.66&  1.74 & +0.55  & +0.79  & +0.85 \\
B3-f3& 95424C3 & 4400 & 1.9 & 0.06 & 1.7  & 4637& 2.96& 0.24 &  1.89  & +0.23 & +0.31  & +0.12  \\
B3-f4& 208959C6 & 4400 & 2.1 & 0.09 & 1.5 & 4319& 2.60& $-$0.12&  1.50 & +0.33  & +0.59   & +0.55  \\
B3-f7& 279577C7 &  4800 & 2.1 & 0.16 & 1.7& 4517& 2.93& 0.17 &  1.55 & +0.44 & +0.30  & +0.15  \\
B3-f8& 193190C5 & 4800 & 1.9 & 0.20 & 1.5 & 4436& 2.88& 0.24 &  1.54 & +0.61   & +0.25   & +0.27  \\
BW-b2& 214192 & 4300 & 1.9 & 0.22 & 1.5   & 4367& 2.39& 0.18 &  1.68 & +0.10 & +0.28 & +0.19   \\
BW-b5& 82760 &  4000 & 1.6 & 0.17 & 1.2   & 3939& 1.68& 0.25 & 1.31 & +0.14 & $-$0.10   & $-$0.10   \\
BW-b6& 392931 & 4200 & 1.7 & $-$0.25 & 1.3  & 4262& 1.98& $-$0.32&  1.44 & +0.50 & +0.55 & +0.49   \\
BW-f1& 433669 & 4400 & 1.8 & 0.32 & 1.6   & 4359& 2.51& 0.28 &  1.93& +0.85  & +0.50  & +0.37   \\
BW-f5& 240260 & 4800 & 1.9 & $-$0.59 & 1.3  & 4818& 2.89& $-$0.51&  1.29 & +0.11  & +0.26  & +0.31  \\
BW-f6& 392918 & 4100 & 1.7 & $-$0.21 & 1.5  & 4117& 1.43& $-$0.43&  1.69 & +0.22  & +0.52   & +0.42 \\
B6-b1& 29280c3 & 4400 & 1.8 & 0.07 & 1.6  & 4372& 2.59& 0.25 &  1.57 & +0.40  & +0.17 & +0.22\\
B6-b3& 31220c2 & 4700 & 2.0 & 0.10 & 1.6  &     4468& 2.48& 0.05 &  1.67 & +0.25 & +0.15 & +0.04 \\
B6-b4& 60208c7 & 4400 & 1.9 & $-$0.41 & 1.7 &   4215& 1.38& $-$0.62&  1.68 & +0.30  & +0.43   & +0.33  \\
B6-b5& 31090c2 & 4600 & 1.9 & $-$0.37 & 1.3 &   4340& 2.02& $-$0.48&  1.34 & +0.36 & +0.49 & +0.45 \\
B6-b6& 77743c7 & 4600 & 1.9 & 0.11 & 1.8  &     4396& 2.37& 0.19 &  1.77 & +0.52  & +0.25  & +0.22    \\
B6-b8& 108051c7 & 4100 & 1.6 & 0.03 & 1.3 & 4021& 1.90& 0.06 & 1.45 & +0.22  & +0.19  & +0.21   \\
B6-f1& 23017c3 & 4200 & 1.6 & $-$0.01 & 1.5 &   4149& 2.01& 0.10 & 1.65 & +0.08  & +0.15  & $-$0.10   \\
B6-f3& 21259c2 & 4800 & 1.9 & $-$0.29 & 1.3 &   4565& 2.60& $-$0.35& 1.28 & +0.26 & +0.42  & +0.41   \\
B6-f5& 33058c2 & 4500 & 1.8 & $-$0.37 & 1.4 &   4345& 2.32& $-$0.33 & 1.41 & +0.15  & +0.41 & +0.39 \\
B6-f7& 100047c6& 4300 & 1.7 & $-$0.42 & 1.6 & 4250& 2.10& $-$0.31& 1.65 & +0.09  & +0.25  & +0.26 \\
B6-f8& 11653c3 & 4900 & 1.8 & 0.04 & 1.6  & 4470& 2.78& +0.13 & 1.30 & +0.32  & +0.25 & +0.18 \\
BL-1 & 1458c3 & 4500 & 2.1 & $-$0.16 & 1.5  &   4370& 2.19& $-$0.19& 1.50 & +0.20  & +0.35  & +0.35 \\
BL-3 & 1859c2 & 4500 & 2.3 & $-$0.03 & 1.4  &   4555& 2.48& $-$0.09& 1.53 & +0.25 & +0.18 & +0.30 \\
BL-4 & 3328c6 & 4700 & 2.0 & 0.13 & 1.5   & 4476& 2.94& 0.27 & 1.41 & +0.43  & $-$0.14  & +0.16 \\
BL-5 & 1932c2 & 4500 & 2.1 & 0.16 & 1.6   & 4425& 2.65& 0.28 & 1.68 & +0.13  & +0.00 & +0.02  \\
BL-7 & 6336c7 & 4700 & 2.4 & $-$0.47 & 1.4  &   4776& 2.52& $-$0.50 & 1.53 & +0.10  & +0.40 & +0.35 \\
\hline
\hline
\end{tabular}
\end{table*}

\section{Atomic constants}

Table \ref{linelist} reports the lines 
and constants of
Line list of \ion{Na}{I} \citep{froese02}, 
\ion{Mg}{I} \citep[see][]{butler93}. \ion{Al}{I} \citep[see][]{mendoza95}.
 analysed in this work.
Oscillator strengths = log gf values from
NIST (Kramida et al. 2024)\footnote{https://physics.nist.gov/asd}, or APOGEE \citep{smith21}, Kurucz \citep{kurucz95} and VALD3 \citep{ryabchikova15} are listed, showing some differences from these
3 main sources. 
The adopted values are those from NIST
or VALD3 in the optical and APOGEE in the H band.
In column Notes (TURBO) are indicated the lines
for which hyperfine structure is taken into account.


\begin{table*}
\small
\caption{Line list of \ion{Na}{I}, \ion{Mg}{I}, and \ion{Al {I}, C$^2$, CN, and [OI]}. 
}
\label{linelist}
\begin{flushleft}
\begin{tabular}{lcccccccccccccc}
\hline\hline
\noalign{\smallskip}
\hbox{Species} & \hbox{$\lambda$} & \hbox{$\chi_{ex}$}  &\hbox{log~gf} &\hbox{log~gf} &\hbox{log~gf} & Notes & \\
& \hbox{(\AA)} &\hbox{(eV)} & \hbox{(NIST/APOGEE)} & \hbox{(Kurucz)} & \hbox{(VALD)} & \hbox{(TURBO )}  &   \\ 
\noalign{\smallskip}
\hline
\noalign{\smallskip}
\multicolumn{7}{c}{ Optical}  \\ 
\noalign{\smallskip}
\hline
\noalign{\smallskip}
\hbox{NaI} & 5682.6333 & 2.102 & $-$0.706  & $-$0.700 & $-$0.706 & hfs & \\
& 5688.1934 & 2.104 & $-$1.406 & $-$1.400 & $-$1.406 & hfs &  \\
& 5688.2046 & 2.104 & $-$0.452  & $-$0.450 & $-$0.452 & hfs & \\
& 6154.2253 & 2.102 & $-$1.547  & $-$1.560 & $-$1.547 & hfs & \\
& 6160.7470 & 2.104 & $-$1.246  & $-$1.260 & $-$1.246 & hfs & \\
\hbox{MgI} & 6318.716 & 5.108  & $-$2.103 & $-$1.950 & $-$2.103 & $-$2.103 & \\
 & 6319.236 & 5.108 & $-$2.324 & $-$1.730 & $-$2.324 &  $-$2.324 & \\
& 6319.493 & 5.108  & $-$2.803 & $-$2.430 & $-$2.803 & $-$2.803 & \\
\hbox{AlI} & 6696.015 & 3.143 & $-$1.569 & $-$1.347 & $-$1.460 & hfs & \\ 
 & 6698.673 & 3.143  & $-$1.870 & $-$1.647 & $-$1.760 & hfs & \\
 \hbox{C$_2$(0,1)} &  5635.5 & --- & --- & --- & --- &  & \\
 \hbox{CN(5,1)} &  6332.2 & --- & --- & --- & --- &  & \\
 \hbox{[OI]} & 6300.304 & 0.0 & -9.819 & --- & --- &  & \\
 \noalign{\smallskip}
\hline
\noalign{\smallskip}
\multicolumn{7}{c}{ H band}  \\ 
\noalign{\smallskip}
\hline
\noalign{\smallskip}
\hbox{NaI} 
& 16388.858 & 3.754 & $-$1.030 & $-$1.030 & $-$1.027 & hfs& \\
\hbox{AlI} & 16718.957 & 4.085 & 0.290 & 0.152 & 0.220 & hfs&   \\
& 16750.539 & 4.088 & 0.408 & 0.408 & 0.408 &  hfs &   \\
& 16763.359 & 4.087 & $-$0.524 & $-$0.550 & $-$0.480  & hfs &   \\
\noalign{\hrule\vskip 0.1cm} 
\hline                  
\end{tabular}
\end{flushleft}
\end{table*}

\section{Hyperfine structure constants for \ion{Na}{I} lines}

Table \ref{hfs} shows the  atomic hyperfine constants
used to compute the hyperfine structure of the 
\ion{Na}{I} 5682.6333, 5688.1934, and 5688.2046 {\rm \AA} lines:
the dipolar constant A, and the quadrupole coupling constant B,
with the constants of Na P$_{1/2,3/2}$ by \citet{wijngaarden94} 
and and excellent report bringing all the constants by \citet{allegrini22}.
Sodium has the unique species $^{23}$Na \citep{asplund09}.
The HFS was taken into account by applying the code made available 
by \citet{mcwilliam13}.

The hyperfine structure for lines 
\ion{Na}{I} 6154 and 6160 {\rm \AA} and
\ion{Al}{I} 6696 and 6698 {\rm \AA} are already
incorporated and available
in the line list from VALD3.

\begin{table*}
\begin{flushleft}
\scalefont{0.8}
\caption{Atomic constants for \ion{Na}{I} used to compute hyperfine structure:
A and B constants from   \citet{wijngaarden94} and     \citet{allegrini22}.
}             
\label{hfs}      
\centering          
\begin{tabular}{lc@{}c@{}c@{}c@{}c@{}c@{}c@{}c@{}c@{}c@{}c@{}c@{}ccc} 
\noalign{\smallskip}
\hline\hline    
\noalign{\smallskip}
\noalign{\vskip 0.1cm} 
Species & $\lambda$ ({\rm \AA}) & \phantom{-}Lower level 
& \phantom{-}J & \phantom{-}A(MHz) 
& \phantom{-}B(MHz) & \phantom{-}Upper level  
& \phantom{-}J & 
\phantom{-}A(MHz) & \phantom{-}B(MHz)  \\
\noalign{\vskip 0.1cm}
\noalign{\hrule\vskip 0.1cm}
\noalign{\vskip 0.1cm}

$^{23}$NaI & 5682.6333 & 3p 2P & 0.5  & 94.3  &  0.0 &  4d 2D  & 1.5 &  0.527 &  0.0  &   \\

$^{23}$NaI & 5688.1934 & 3p 2P  & 1.5 & 18.64  &  2.77 & 4d 2D  & 1.5 &  0.527 &  0.0  &   \\

$^{23}$NaI & 5688.2046 & 3p 2P  & 1.5 & 18.64  &  2.77 & 4d 2D  & 2.5  &  0.1085 &  0.0  &   \\

\noalign{\vskip 0.1cm}
\noalign{\hrule\vskip 0.1cm}
\noalign{\vskip 0.1cm}  
\hline    
              
\end{tabular}
\tablefoot{Constants are given in MHz.}
\end{flushleft}
\end{table*} 

\end{appendix}
\end{document}